\documentclass[aps,twocolumn,superscriptaddress,preprintnumbers,groupedaddress]{revtex4}  
\usepackage{graphicx}  
\usepackage{dcolumn}   
\usepackage{bm}        
\usepackage{amssymb}   
\usepackage{amsmath,amssymb,bm,graphicx,bbold,epsf,colordvi}
    \usepackage{dsfont}
\usepackage{color}
\newcommand{\lsim}{\buildrel < \over {_\sim}}

\newcommand{\be}{\begin{equation}}
\newcommand{\ee}{\end{equation}}

\def\OMIT#1{{}}

\newcommand{\mcdot}{\!\cdot\!}

\newcommand{\e}{\mathrm{e}}

\newcommand{\eq}[1]{Eq.~\eqref{#1}}

\newcommand{\CPV}{\text{CP}\!\!\!\!\!\!\!\raisebox{0pt}{\small$\diagup$}}

\newcommand{\diag}{\mathrm{diag}}

\begin{document}
\preprint{\vbox{\hbox{ACFI-T15-12}}}
\widetext


\title{Two-Step Electroweak Baryogenesis}
%
%
\author{Satoru Inoue} \affiliation{Physics Department, University of Massachusetts Amherst, Amherst, MA 01003, USA}\affiliation{Department of Physics and Astronomy, University of South Carolina, Columbia, SC 29208, USA}
\author{Grigory Ovanesyan} \affiliation{Physics Department, University of Massachusetts Amherst, Amherst, MA 01003, USA}
\author{Michael J. Ramsey-Musolf} \affiliation{Physics Department, University of Massachusetts Amherst, Amherst, MA 01003, USA}\affiliation{Kellogg Radiation Laboratory, California Institute of Technology, Pasadena, CA 91125, USA}

%
%
%
\vskip 0.25cm

\date{\today}

\begin{abstract}
We analyze electroweak baryogenesis during a two-step electroweak symmetry breaking transition, wherein the baryon asymmetry is generated during the first step and preserved during the second. Focusing on the dynamics of CP-violation required for asymmetry generation, we discuss general considerations for successful two-step baryogenesis. Using a concrete model realization, we illustrate in detail the viability of this scenario and the implications for present and future electric dipole moment (EDM) searches. We find that CP-violation associated with a partially excluded sector may yield the observed baryon asymmetry while evading present and future EDM constraints.

\end{abstract}

\maketitle

\section{Introduction}

The origin of the cosmic matter-antimatter asymmetry remains one of the outstanding mysteries at the interface of particle and nuclear physics with cosmology. The asymmetry is typically characterized by the baryon-to-photon ratio
\be
Y_B \equiv \rho_B /s=(8.59\pm0.11)\times 10^{-11}
\ee
where $\rho_B$ and $s$ are the baryon number and entropy densities, respectively, and where the value has been obtained using data from Planck~\cite{Ade:2013zuv},  WMAP~\cite{Komatsu:2010fb} and  large scale structure measurements.  Assuming a matter-antimatter symmetric Universe at the end of the inflationary epoch, a non-vanishing $Y_B$ can be generated if the microphysics of the early Universe satisfies the well-known three \lq\lq Sakharov conditions"~\cite{Sakharov:1967dj}: (1) baryon number violation; (2) C and CP symmetry violation; (3) departure from thermal equilibrium (or violation of CPT invariance). 

A variety of scenarios that satisfy these criteria have been proposed, each corresponding to a different epoch in cosmic history. One of the most widely considered is electroweak baryogenesis (EWBG), wherein the baryon asymmetry, $Y_B$, was generated during the era of electroweak symmetry-breaking (EWSB) that occurred roughly ten picoseconds after the Big Bang. Successful EWBG requires a strong first order electroweak phase transition (EWPT) that proceeds via bubble nucleation. CP-violating asymmetries generated at the bubble walls diffuse ahead of the expanding bubbles, catalyzing baryon number creation through electroweak (EW) sphalerons. The expanding bubbles capture the non-vanishing baryon number, which is preserved in the bubble interiors if the transition is sufficiently strong so as to quench the electroweak sphalerons. For a recent review of this scenario, see Ref.\cite{Morrissey:2012db} and the references therein.

In principle, the Standard Model (SM) contains all the elements needed to satisfy the Sakharov criteria in the context of EWBG. In practice, the value of the Higgs boson mass is too large to accommodate a first order EWPT; in a SM Universe, the EWSB transition is of a cross-over type\cite{Gurtler:1997hr,Laine:1998jb,Csikor:1998eu,Aoki:1999fi}. Even if the Higgs boson had been sufficiently light, the CP-violating asymmetries associated with CP-violating phase of the Cabibbo-Kobayashi-Maskawa matrix are too small to lead to the observed value of $Y_B$\cite{Gavela:1993ts,Huet:1994jb,Gavela:1994dt}. Thus, successful EWBG requires both new degrees of freedom to induce a first order EWPT and new CP-violating interactions to generate sufficiently sizable asymmetries during the transition. 

In this study, we consider EWBG in light of the possibility that EWSB proceeded in multiple steps, rather than in a single transition from an electroweak-symmetric to an EWSB vacuum as has been conventionally assumed. We further consider the possibility that $Y_B$ was generated during a transition to a EWSB-vacuum that is not the present day vacuum (\lq\lq Higgs phase"), but rather one that was the lowest energy state for a period prior to the final transition to the Higgs phase of the SM. For simplicity, we concentrate on a two-step scenario, though the general features could generalize to patterns of EWSB that entail additional intermediate phases. 

In earlier work, we demonstrated the viability of the two-step EWSB scenario induced by the presence of an electroweak triplet $\Sigma\sim (1,3,0)$ \cite{Patel:2012pi}, in which the transition to the first EWSB-vacuum involving only a non-vanishing neutral triplet vacuum expectation value (vev) is strongly first order. Some general conditions needed for a successful two-step EWPT in extensions of the SM Higgs sector were subsequently studied in Ref.\cite{Blinov:2015sna}. Here, we focus on the generation of CP-violating asymmetries during the transition to the penultimate EWSB-vacuum. Starting with $\Sigma$-extended SM, we show that generation of these asymmetries requires additional field content. After discussing general considerations, we focus on a concrete example that provides a proof-in-principle of viability of the general paradigm. 

It is interesting to ask about the experimental signatures of the multi-step scenario. Requiring that the final transition to the SM Higgs phase occurs at sufficiently low temperatures as to avoid baryon number erasure through re-excited EW sphalerons implies that at least a subset of the mass parameters in the Lagrangian are not too different from the EW scale. Initial studies of the consequences for collider phenomenology in the $\Sigma$-extended SM are discussed in Refs.~\cite{Patel:2012pi,FileviezPerez:2008bj}. In general, the introduction of new CP-violating interactions must contend with severe constraints from searches for the permanent electric dipole moments (EDMs) of atoms, molecules, and the neutron (see, {\em e.g.} Refs.~\cite{Chupp:2014gka,Engel:2013lsa,Pospelov:2005pr}) as well as possibly probes of CP-violation (CPV) in the heavy flavor sector for some scenarios\cite{Liu:2011jh,Tulin:2011wi,Cline:2011mm}. It is reasonable to expect that any new CPV as needed for successful EWBG will be testable with the next generation EDM searches or heavy flavor studies. In what follows, we show that this expectation may not be borne out for the multi-step scenario, as the CP-violating interactions may involve a partially secluded sector whose impacts on low-energy CP-violating observables are highly suppressed. In short, multi-step electroweak baryogenesis may open a new window for generation of $Y_B$ at the weak scale, one that is relatively immune to experimental constraints on CPV in the near term. 

Before proceeding, we note that others have considered baryogenesis scenarios going beyond the conventional paradigm of a one-step EWSB.  Ref.~\cite{Land:1992sm} considered a two-step phase transition (2SPT) scenario using a Two Higgs Doublet Model (2HDM), wherein  the first step is a second order (or cross over) PT, while the second step is strongly first order. However, because electroweak symmetry is broken during the first step, the $B+L$ violating processes are suppressed, and the CPV asymmetries generated during the second step cannot be efficiently transferred into the baryon asymmetry\cite{Hammerschmitt:1994fn}. The authors of Ref.~\cite{Jiang:2015cwa} considered an extension of the SM with a complex scalar singlet\cite{Barger:2008jx}, in which the universe first undergoes a transition to an EW-symmetric phase with a non-vanishing singlet vev, followed by a transition to the Higgs phase. CPV-asymmetries are induced during the second step by a non-renormalizable Higgs-singlet-top quark interaction. Our scenario differs qualitatively from these earlier studies, since $Y_B$ is generated during the first EWSB transition and is preserved during the subsequent transition to the Higgs phase. In this initial study, we also focus solely on CPV in the scalar sector involving only renormalizable operators.


Our discussion of two-step electroweak baryogenesis is organized as follows. In section \ref{sec:general}, we outline general considerations for baryogenesis in this scenario. In section \ref{sec:model} we define the details of our model with the particle content, interactions, relevant Feynman rules and the conditions for EWSB. Section \ref{sec:pheno}  gives the framework for implementing constraints from two relevant observables: EDMs and Higgs to diphoton decay rate. In section \ref{sec:transport} we present the set of transport equations that describe the dynamics of particle-antiparticle asymmetry generation during the first step of the two-step transition. Finally we present our numerical results in section \ref{sec:results} and conclude in section \ref{sec:conclude}. Technical details associated with solving the transport equations appear in Appendices. A reader interested primarily in the general framework, specific model realization, and primary results may wish to concentrate on section \ref{sec:general}, the first part of section \ref{sec:model}, and Figs.~\ref{fig:ModelsABmh2},\ref{fig:ModelsABtanbeta} of section \ref{sec:results} that show the sensitivity of present and future EDM searches to regions of the model parameter space consistent with the observed baryon asymmetry.

\section{General Considerations }
\label{sec:general}
In what follows, we will adopt a specific model realization to illustrate the viability of two-step electroweak baryogenesis. Before doing so, we provide some general considerations that should guide the choice of a model, concentrating here on the ingredients needed to generate a baryon asymmetry during the first step. We note that generation and preservation of the baryon asymmetry during the first step requires that it be a strong first order EWSB transition, while the preservation of this asymmetry during the second step to the SM Higgs phase requires that (a) the temperature of the latter transition is sufficiently low as to avoid re-exciting the electroweak sphalerons and (b) the entropy released during the second transition be sufficiently small so as to avoid over dilution of the asymmetry generated during the first step. The possibility of satisfying these requirements was demonstrated in our previous work on the $\Sigma$-extended SM \cite{Patel:2012pi}.

In order to produce a non-vanishing $Y_B$ during the first transition, one requires generation of CP-violating asymmetries that ultimately yield a non-vanishing number density of left-handed (LH)  fermions, $n_L$. The latter biases the electroweak sphalerons in the unbroken phase ahead of the advancing bubble walls whose interiors contain the first broken phase. We consider two sectors for this purpose: (a) the SM and (b) a new sector that contains the fields responsible for the first EWSB transition -- generically denoted $\phi_j$ --  plus additional fields that interact with these fields and that may be partially or completely secluded from the SM. The following possibilities then emerge:
\begin{itemize}
\item The new sector contains additional LH fermions that contribute to the B+L anomaly. CP-violating interactions of these fermions with the $\phi_j$ lead to a non-vanishing $n_L$.
\item A CP-violating asymmetry is generated for one or more of the new sector scalar fields $\phi_j$ and is subsequently transferred to the LH fermions of the SM through interactions between the two sectors.
\item A CP-violating asymmetry involving SM fields is generated through their interactions with the $\phi_j$ yielding a non-vanishing $n_L$ either directly or indirectly via SM Yukawa and gauge interactions.
\end{itemize}

In the remainder of this initial study, we concentrate on the third possibility, as it presents the greatest potential for experimental accessibility; we defer a consideration of the other possibilities for future work. We also focus on the case where the initial CP-violating asymmetry is generated entirely in the scalar sector and transferred to the SM LH fermions via Yukawa interactions. In this instance, one requires at least two distinct scalar fields that mix through their interactions with the bubble walls of the first EWSB transition. For concreteness, we will utilize a 2HDM for the Higgs sector, where the doublet fields mix during the first EWSB transition. During this first step, however, neither of the doublets obtain a vacuum expectation value (vev). Consequently, we may treat both the neutral and charged components of the doublets as complex scalars with masses determined by the finite-temperature potential and the space-time varying vevs of the fields driving the first EWSB transition. This treatment differs from what is appropriate when the neutral components of the doublets obtain vevs, leading to one combination of the CP-odd neutral scalar that is eaten by the $Z$-boson to become its longitudinal component and the other combination that is a physical CP-odd scalar. The latter framework, wherein the neutral CP-even and CP-odd scalars are treated as distinct degrees of freedom, applies to the second EWSB vacuum, or Higgs phase.

For the first EWSB step, then, we consider two complex scalars $h_1$ and $h_2$ whose mass-squared matrix has the form
\be
{\mathbf{M}^2 }= \left(
\begin{array}{cc}
m_{11}^2 & m_{12}^2\\
m_{12}^{2\ast} & m_{22}^2
\end{array}
\right)
\label{eq:mixing}
\ee

As discussed in Refs.~\cite{Cirigliano:2009yt,Cirigliano:2011di}, generation of non-vanishing $h_1$ and $h_2$ asymmetries requires that $m_{12}^2$ contain a spacetime-dependent complex phase, $\theta(x)=\mathrm{Arg}(m_{12}^2)$. The latter arises from the interactions of the $h_{1,2}$ with the expanding bubble wall that effectively provides a spacetime-dependent background field {\em via} the space-time variation of the vevs of the fields $\phi_j$ involved in the first EWSB transition. 

Additional comments regarding the origins of $\theta(x)$ are in order. In principle, this phase could arise entirely spontaneously due to the vevs of the $\phi_j$ without any dependence on an explicit CP-violating phase in the Lagrangian of the theory. In practice, the net asymmetry produced by a coalescing ensemble of bubbles admitting only spontaneously-generated CP-violating phases will be zero\cite{Funakubo:1996dw}. For every bubble having a spontaneous phase of a given magnitude and sign, there will always be a partner bubble somewhere in the ensemble with a phase of the same magnitude but opposite sign. The contributions of the two bubbles will then cancel after the bubbles coalesce and the transition completes. Breaking the energy degeneracy between would-be partner bubbles requires the presence of an explicit CP-violating phase in the theory\cite{Comelli:1993ne}.

Under these conditions, $\theta(x)$ will be space-time dependent only when two distinct fields with differing non-constant space-time profiles contribute to $m_{12}^2$, even in the presence of an explicit phase in the scalar potential. For the two-step transition of interest here, wherein the doublet vevs remain zero during the first step, possible forms of the CPV interactions that satisfy these considerations include

\noindent (a) CPV-asymmetries generated entirely in the new scalar sector:
\be
\phi_1^\dag \phi_2 \phi_j^\dag \phi_j +\mathrm{h.c.}\ \ \ , \qquad j=1,2
\ee

\noindent (b) CPV-asymmetries generated in the SM (doublet) scalar sector:
\be
\label{eq:portal}
\phi_j^\dag \phi_j H_1^\dag H_2 + \mathrm{h.c}, \qquad \phi_j^\dag \phi_k H_1^\dag H_1+\mathrm{h.c.}\ \ \ , \mathrm{etc.}
\ee

Note that the coefficients of these operators may be complex, yielding the requisite explicit CP-violating phase. Note also that only one of the $\phi_j$ need carry SM electroweak charges, as required for quenching of the electroweak sphalerons during the first EWSB transition. The remaining new sector scalar fields may be pure gauge singlets or charged under other symmetries that do not contain the SM electroweak symmetries as a subgroup.

\section{The Model}
\label{sec:model}
\begin{figure}
\includegraphics[scale=0.35]{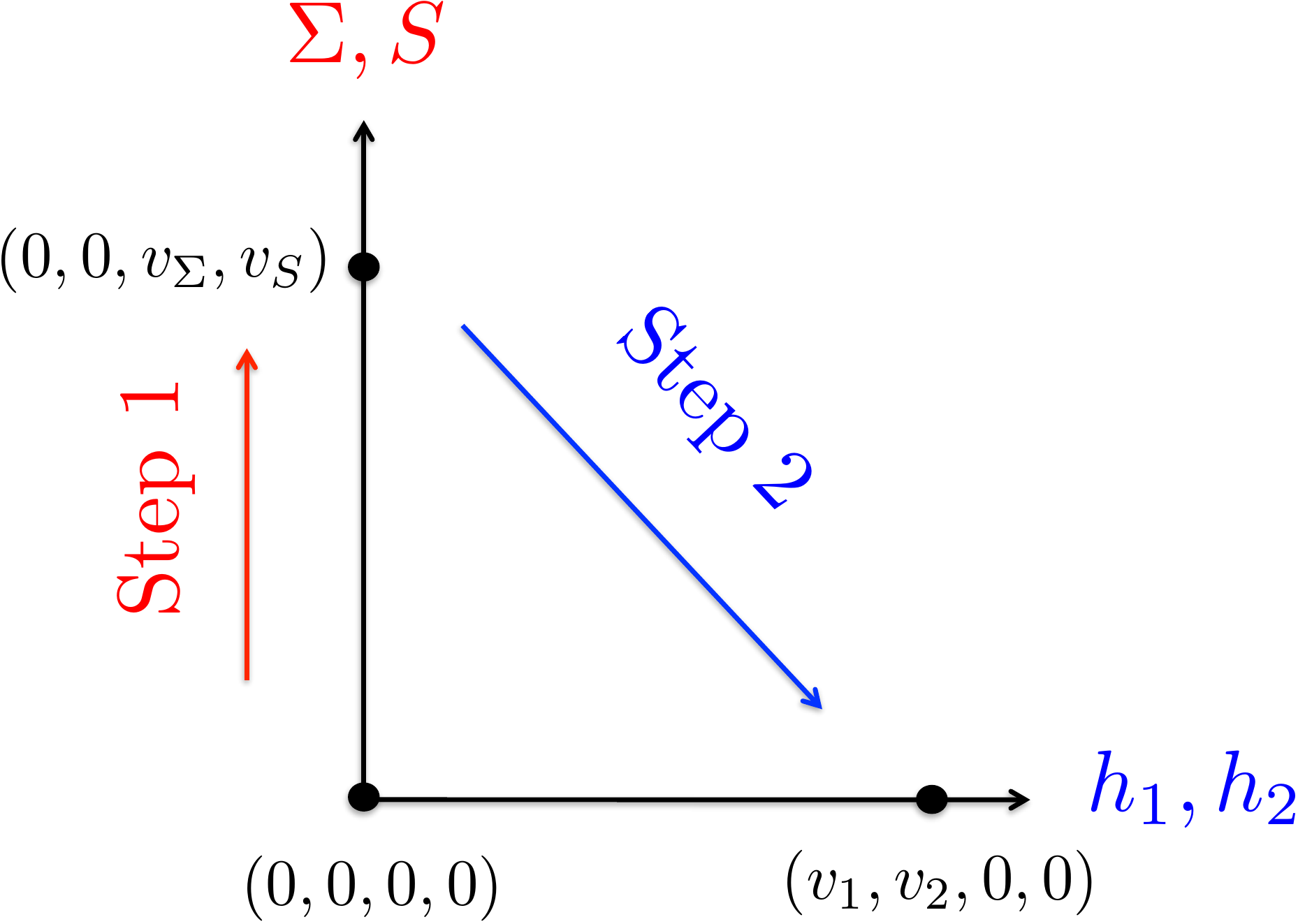}
\caption{\label{fig:TwoStepIllustration} Illustration of the two-step PT that we study in this paper. We focus on the BAU generated during the first step.}
\end{figure}
With the foregoing considerations in mind, we illustrate the viability of two-step EWBG with a concrete model. As indicated above, for the SM sector we extend the theory to a 2HDM in order to allow for mixing between two SM-sector states. For the new sector we choose $\phi_1$ to be the real triplet $\Sigma\sim (1,3,0)$ and take the second field $\phi_2$ to be a real singlet $S$. In principle, we could have chosen $\phi_2$ to be a non-singlet with respect to SM symmetries. Our rationale for choosing a singlet is that (a) the dynamics of EWSB during the first step are relatively simple; (b) it is relatively straightforward to write down a potential for $S$ wherein the singlet-vev is non-zero and varying with temperature during the first transition; (c) the presence of the singlet illustrates the phenomenological features associated with a  sector that interacts with the SM sector only through Higgs portal interactions of the type in Eq.~(\ref{eq:portal}).  With these choices, we take the vevs of the singlet and neutral component of the triplet to be, $v_S$ and $v_\Sigma$, respectively. The corresponding potential is


\begin{eqnarray}
&& V(H_1,H_2,\Sigma, S)\nonumber\\
&&=-\frac{\mu_{\Sigma}^2}{2}\left(\vec{\Sigma}\mcdot \vec{\Sigma}\right)+\frac{b_{4\Sigma}}{4}\left(\vec{\Sigma}\mcdot \vec{\Sigma}\right)^2+\frac{b_{2S}}{2}S^2+\frac{b_{4S}}{4}S^4\nonumber\\
&&+\left[\frac{1}{2}\,a_{2\Sigma}\,H_1^{\dagger}H_2\left(\vec{\Sigma}\mcdot \vec{\Sigma}\right)+\frac{1}{2}a_{2S}H_1^{\dagger}H_2 S^2+\text{h.c.}\right]\,,\nonumber\\
&&+ a_{1\Sigma S}\vec{\Sigma}\mcdot \vec{\Sigma}\,S+ \frac{1}{2} a_{2\Sigma S} \vec{\Sigma}\mcdot \vec{\Sigma}\,S^2+V(H_1,H_2)\,.
 \label{eq:deltaV}
\end{eqnarray}
Note that the quartic couplings $a_{2\Sigma}$ and $a_{2S}$ are in general complex. The real couplings $a_{1\Sigma S}$ and $a_{2\Sigma S}$ have little bearing on the dynamics of CPV, whereas they will play a role in the EWPT. For purposes of keeping the number of parameters manageable for this initial study, we will set them to zero.

For the 2HDM potential we consider the $Z_2$-symmetric limit under which $H_1$ and $H_2$ are oppositely charged, thereby alleviates the possibility of potentially dangerous tree-level flavor changing neutral currents (FCNCs): 
\begin{eqnarray}
&&V(H_1,H_2)=\frac{\lambda_1}{2}\left(H_1^{\dagger}H_1\right)^2+\frac{\lambda_2}{2}\left(H_2^{\dagger}H_2\right)^2\nonumber\\
&&+\lambda_3\left(H_1^{\dagger} H_1\right)\left(H_2^{\dagger} H_2\right)+\lambda_4\left(H_1^{\dagger} H_2\right)\left(H_2^{\dagger} H_1\right)\nonumber\\
&&+\frac{1}{2}\left[\lambda_5\left(H_1^{\dagger} H_2\right)^2+\text{h.c.}\right]-\frac{1}{2}\Big\{m_{11}^2\left(H_1^{\dagger}H_1\right)\nonumber\\
&&+m_{22}^2\left(H_2^{\dagger}H_2\right)\Big\}\,.\nonumber\\ \label{eq:twohiggsdoubletlagrangian}
\end{eqnarray}
Note that the $Z_2$-breaking operators $H_1^\dag H_2 \vec{\Sigma}\mcdot \vec{\Sigma}$ and $H_1^{\dagger}H_2 S^2$ generate divergent contributions to the operator $H_1^\dag H_2$ at one-loop order, implying the need for a counter term $m_{12}^2 H_1^\dag H_2+\mathrm{h.c.}$. We retain freedom to choose the finite part of $m_{12}^2$ such that its sum with the finite parts of the one-loop graphs is sufficiently small as to satisfy experimental FCNC bounds. As we discuss in Section \ref{sec:results}, there will also be finite-temperature contributions to $H_1^\dag H_2$ associated with $H_1^\dag H_2 \vec{\Sigma}\mcdot \vec{\Sigma}$ and $H_1^{\dagger}H_2 S^2$ that cannot in general be eliminated and that will lead to mixing between the two doublets in the early universe.

The pattern of two-step EWSB arising from this potential is illustrated in Fig.~\ref{fig:TwoStepIllustration}. At high temperatures, the universe starts in the completely symmetric phase with all vevs set to zero\footnote{An interesting possibility, which we do not explore here, is that the transition proceeds first to a vacuum having only a non-vanishing singlet vev}. At a critical temperature $T_\Sigma$, the vacuum with non-vanishing $v_\Sigma$ and $v_S$ becomes degenerate in energy with the symmetric phase; just below $T_\Sigma$, the first order transition to the \lq\lq $\Sigma$ phase" proceeds through bubble nucleation. This transition is denoted by \lq\lq Step 1" in Fig.~\ref{fig:TwoStepIllustration}. At a lower temperature $T_H$, the vacuum with non-vanishing $v_1$ and $v_2$ becomes degenerate with the $\Sigma$ phase, and as the universe cools further, a transition to the Higgs phase occurs (\lq\lq Step 2"). In the $\Sigma$-extended SM, it was observed that the second transition is also typically first order. However, EWBG is not viable in this step because the sphalerons that are active before the first transition have been quenched in the $\Sigma$ phase\footnote{The $\Sigma$ phase does admit a baryon number violating monopole solution, but the requirements on the parameters of the theory need to ensure baryon number preservation appear to be commensurate with those relevant to baryon number preservation in the Higgs phase where there is a sphaleron solution.}. 

We will then investigate the possibility of EWBG during the first step. As a preamble, we first analyze the structure of the potential in greater detail. After EWSB occurs in the second step, the neutral components of each of the Higgs fields acquire a vev, and the fluctuations around this value can be characterized by charged ($H_{i}^+$), CP even ($H_{i}^0$) and CP odd $(A_{i}^0)$ fields respectively:
\begin{eqnarray}
H_i=\left(\begin{array}{cc}
H_i^+\\
\frac{v_i+H_{i}^0+i A_{i}^0}{\sqrt{2}}
\\
\end{array}\right),\qquad \text{where\,\,\,\,} i=1,2\,.
\end{eqnarray}

The  potential $V(H_1,H_2,\Sigma, S)$ has three complex couplings: $\lambda_5,\,\,a_{2\Sigma},\,a_{2S}$. An overall phase in these couplings is unphysical and can be rotated away via a rephasing transformation on the complex scalar fields
\begin{eqnarray}
H_1= \e^{i\theta_1}H_1'\,,\quad H_2=\e^{i\theta_2}H_2'\,, \quad \Sigma=\Sigma'\,,\quad S=S'\,.\label{eq:rephasing1}
\end{eqnarray}
A global phase $\theta_2-\theta_1$ can be absorbed into the following redefinition of the couplings and vevs:
\begin{eqnarray}
&& \lambda_5'=\e^{2i(\theta_2-\theta_1)}\lambda_5\,,\qquad\,\,\,\,\,\,\,\,\,\,\, \left(v_1v_2^*\right)' = \e^{i(\theta_2-\theta_1)}v_1 v_2^*\nonumber\\
&&a_{2\Sigma}'=\e^{i(\theta_2-\theta_1)}a_{2\Sigma}\,, \qquad\,\,\,\,\,\,\,\,\, a_{2S}'=\e^{i(\theta_2-\theta_1)}a_{2S}\,,\nonumber\\\label{eq:phasereparametrization}
\end{eqnarray}
Without loss of generality, we assume $v_1=v_1^*$ and $v_2=|v_2|\e^{i\xi}$. The second equation in the first line leads to $\xi'=\xi+\theta_1-\theta_2$ rephasing transformation on the spontaneously generated phase. The transformation in \eq{eq:rephasing1} and \eq{eq:phasereparametrization} leaves the Lagrangian unchanged and, thus, phases in the couplings that can be eliminated with such redefinition are unphysical. The model then contains the following three physical phases:
\begin{eqnarray}
&&\delta_{\Sigma}=\arg\left[a_{2\Sigma}^*\,v_{1}v_{2}^*\right]\,,\nonumber\\
&&\delta_S=\arg\left[a_{2S}^*\,v_{1}v_{2}^*\right]\,,\nonumber\\
&&\delta_{\lambda_5}=\arg\left[\lambda_5^*\,\left(v_{1}v_2^{*}\right)^2\right]\,.\label{eq:deltaiphases}
\end{eqnarray}
Due to the $\rho$ parameter constraints, at the zero-temperature triplet vev must be small. In what follows, we set it to zero: $\langle\Sigma\rangle=0$. For simplicity, we will also take the zero temperature vev of $S$ to vanish. Given the  symmetry of the  potential under $S\to -S$ and $\Sigma\to -\Sigma$, both $S$ and the neutral triplet may contribute to the dark matter relic density in this case. 

\subsection{Minimizing the potential}
Stability of the vacuum state, after the transition to the second EWSB-vacuum, requires that the Lagrangian couplings and the vevs of the Higgs bosons satisfy the minimization conditions on the potential and positivity of all the masses in the spectrum. For $V(H_1,H_2,\Sigma, S)$ given in  \eq{eq:deltaV} and \eq{eq:twohiggsdoubletlagrangian} we obtain the following minimization conditions:
\begin{eqnarray}
&&m_{11}^2= v^2 \left[\lambda_1 {\cos^2\beta}+(\lambda_3+\lambda_4+|\lambda_5| \cos\delta_{\lambda_5}) {\sin^2\beta}\right]\nonumber\\
&&\,\,\,\qquad\qquad+\left(|a_{2\Sigma}|v_{\Sigma}^2 \cos\delta_{\Sigma} +|a_{2S}|v_{S}^2 \cos\delta_{S}\right)\tan\beta,\nonumber\\
&&m_{22}^2=v^2 \left[\lambda_2\sin^2\beta+ (\lambda_3+\lambda_4+|\lambda_5| \cos\delta_{\lambda_5})\cos^2\beta\right]\nonumber\\
&&\,\,\,\qquad\qquad+\left(|a_{2\Sigma}| v_{\Sigma}^2 \cos\delta_{\Sigma}+|a_{2S}| v_{S}^2 \cos\delta_{S}\right) \cot\beta,\nonumber\\
&&0= \frac{v^2\sin\beta\cos\beta }{2}\Big(\sin\beta\cos\beta \,v^2 |\lambda_5| \sin\delta_{\lambda_5}\nonumber\\
&&\qquad\qquad\qquad\qquad+|a_{2\Sigma}| v_{\Sigma}^2 \sin\delta_{\Sigma}+|a_{2S}| v_{S}^2 \sin\delta_{S}\Big),\nonumber\\
&&0=v_{\Sigma}
\left(b_4 v_{\Sigma}^2-\mu_{\Sigma}^2+|a_{2\Sigma}| v^2 \cos\delta_{\Sigma} \sin\beta\cos\beta\right),\nonumber\\
&&0=v_{S}
\left(b_{2S} +v_Sb_{4S}+|a_{2S}| v^2 \cos\delta_{S} \sin\beta\cos\beta\right),\nonumber\\ \label{eq:fullx0minimizationcoditions}
\end{eqnarray}
where $v\equiv \sqrt{|v_1|^2+|v_2|^2}, \,\tan\beta\equiv |v_2|/|v_1|$. In this paper we concentrate on the case of {\it{no spontaneous CP violation}}, i.e. $\xi=0$\, for simplicity. In this case the three physical phases are related to the complex couplings in the Lagrangian in the following way 
\begin{eqnarray}
\delta_{\Sigma}=-\arg a_{2\Sigma}, \quad\delta_{S}=-\text{arg}\,a_{2S}, \quad\delta_{\lambda_5}=-\text{arg}\,\lambda_5\,.
\end{eqnarray}
Note, that the phases $\delta$ are manifestly rephasing invariant, while the arguments of the complex couplings  $\lambda_5, a_{2\Sigma}, a_{2S}$ are not. The expressions above apply in the rephasing basis corresponding to the $\xi=0$ choice that we have made.

In the limit $v_{\Sigma}=v_{S}=0$ we obtain minimization conditions that are identical to those of the 2HDM \cite{Inoue:2014nva}
\begin{eqnarray}
&&m_{11}^2= v^2 \left[\lambda_1 {c_{\beta}^2}+(\lambda_3+\lambda_4+\text{Re}\, \lambda_5)\, {s_{\beta}^2}\right] ,\nonumber\\
&&m_{22}^2=v^2 \left[\lambda_2s_{\beta}^2+ (\lambda_3+\lambda_4+\text{Re}\,\lambda_5)\,c_{\beta}^2\right]\,,\nonumber\\
&&0= -c_{\beta}s_{\beta} \,v^2 \,\text{Im}\lambda_5\, ,\label{eq:minimizationconditions}
\end{eqnarray}
where $c_{\beta}\equiv\cos\beta$, $s_{\beta}\equiv\sin\beta~$. The fourth and the fifth equations in \eq{eq:fullx0minimizationcoditions}, which correspond to equating to zero the partial derivative with respect to the triplet and singlet vevs correspondingly, become a trivial ``zero equals to zero" equations. Note, however, that prior to the second step of the two-step transition, the latter two equations must be satisfied for non-vanishing $v_\Sigma$ and $v_S$ while $v_1=0=v_2$. 

\subsection{Mass mixing}
We start from the mixing among the charged particles $H_1^+, H_2^+, \Sigma^+$.  The massless Goldstone combination is
\begin{eqnarray}
G^+=c_{\beta}H_1^++s_{\beta}H_2^+\,.
\end{eqnarray}
The remaining two orthogonal charged scalars
\begin{eqnarray}
&&\phi_1^+=-s_{\beta}H_1^++c_{\beta}H_2^+,\nonumber\\
&&\phi_2^+=\Sigma^+,
\end{eqnarray}
do not mix and have a diagonal mass matrix
\begin{eqnarray}
&&M_{\phi_i}^2=\left[ \begin{array}{cc}
m_{H^+}^2&0\\
0  & m_{\Sigma^+}^2 \end{array} \right],
\end{eqnarray}
where the masses are related to the potential parameters via the following equations
\begin{eqnarray}
&&m^2_{H^+}=\frac{1}{2}\left(-\lambda_4-\text{Re}\,\lambda_5\right)v^2,\nonumber\\
&&m_{\Sigma^+}^2=-\mu_{\Sigma}^2+\text{Re}\,a_{2\Sigma}\,v^2\,c_{\beta}s_{\beta}\,.\label{eq:chargedmasses}
\label{eq:massequationsandnudefinition}
\end{eqnarray}
The mass formula for the charged Higgs, $H^+$, agrees with the corresponding $Z_2$ symmetric limit of Ref.\cite{Inoue:2014nva} and is not modified by the presence of additional fields of the triplet and the singlet. 

The neutral scalar bosons $H_1^0, H_2^0, A_1^0, A_2^0, \Sigma^0$ mix and in the most general case ($\xi\ne 0$) there is one massless neutral Goldstone boson $G^0$. In the case of our interest $\xi=0$ this state is the following combination of CP odd Higgs bosons $A_1^0, A_2^0$:
\begin{eqnarray}
G^0=c_{\beta}A_1^0+s_{\beta}A_2^0\,.
\end{eqnarray}
This equation is unchanged compared to the pure 2HDM Ref.\cite{Inoue:2014nva}. Considering the mixing between the orthogonal state $A^0\equiv-s_{\beta}A_1^0+c_{\beta}A_2^0$ and three other neutral scalars: $H_1^0, H_2^0, A^0, \Sigma^0$ we obtain the following mixing matrix
\begin{small}
\begin{eqnarray}
&&M^2_{\text{neutral}}=\nonumber\\
&&v^2\left( \begin{array}{cccc}
\lambda_1c_{\beta}^2 &\lambda_{345}\,s_{\beta}c_{\beta}&0& 0\\
\lambda_{345}\,s_{\beta}c_{\beta} & \lambda_2s_{\beta}^2 &  0& 0 \\
0 & 0 &-\text{Re}\,\lambda_5&0\\
0  & 0 & 0 &\frac{m_{\Sigma^0}^2}{v^2} \end{array} \right),\nonumber\\ \label{eq:neutralmassmixing}
\end{eqnarray}
\end{small}
\noindent where $m_{\Sigma^0}^2\equiv m_{\Sigma^+}^2$\footnote{Note that the charged and neutral triplet masses are split at the one loop level due to EWSB.} and 
\begin{eqnarray}
\lambda_{345}=\lambda_3+\lambda_4+\text{Re}\,\lambda_5.
\end{eqnarray}
 The top left $3\times 3$ block of the matrix $M_{\text{neutral}}^2$ is the same as in the scenario of pure 2HDM \cite{Inoue:2014nva} for $m_{12}^2=0$. Thus, there is no mixing between the triplet and the two Higgs doublets in our theory for both charged and neutral states.
 
 The singlet in our theory does not mix with any other neutral particles and its mass in terms of the parameters of the potential equals
 \begin{eqnarray}
 m_{S}^2=b_{2S}+\text{Re}\,a_{2S}\,v^2\,c_{\beta}s_{\beta}\,.\label{eq:singletmassformula}
 \end{eqnarray} 
 
\subsection{Relevant Feynman rules}
We concentrate on the type-II 2HDM, which is motivated by the minimal supersymmetric extensions of the SM. It has the following interaction Lagrangian between the Higgs bosons and the fermions
\begin{eqnarray}
\mathcal{L}_{\text{II}}^{Y}=-Y_U\,\overline{Q}_L\,i\sigma_2\,H_2^*\,u_R-Y_D\overline{Q}_L\,H_1d_R+\text{h.c.}\,\label{eq:YukawaLagrangian}
\end{eqnarray}
For the EDM constraint we will require the Feynman rules for the Yukawa interactions of the Higgs bosons with the fermions $h_i\bar{f}f$, the tri-scalar interactions $h_i\Sigma^+\Sigma^-$, and the couplings with neutral gauge bosons $Z\,\Sigma^+\Sigma^-,\,\gamma\,\Sigma^+\Sigma^-$. The relevant interaction Lagrangian reads
\begin{eqnarray}
&&\mathcal{L}_{\text{int}}=-\frac{m_f}{v}\,\left[h_i\,\left(c_{f,i}\bar{f}f+\tilde{c}_{f,i}\,\bar{f}i\gamma_5 f\right)+G^0\,\tilde{d}_{f}\,\bar{f}i\gamma_5 f\right]\nonumber\\
&&\,\,\,\,\,\,\qquad-\bar{\lambda}_iv h_i \Sigma^+\Sigma^-\,,\nonumber\\
&&+\left[\Sigma^+\,\left(i\partial_{\mu}\Sigma^-\right)-\left(i\partial_{\mu}\Sigma^+\right)\Sigma^-\right]\left(e\, A_{\mu}+g_2\,c_W\,Z_{\mu}\right)\,,\nonumber\\
\end{eqnarray}
where the couplings $c, \tilde{c}, \bar{\lambda}$ equal to
\begin{eqnarray}
&&c_{t,i}=R_{i2}/s_{\beta},\qquad c_{b,i}=R_{i1}/c_{\beta},\nonumber\\
&&\tilde{c}_{t,i}=-R_{i3}/t_{\beta},\,\,\,\,\,\,\,\, \tilde{c}_{b,i}=-R_{i3}t_{\beta},\label{eq:feynmanrulescouplings}\\
&&\tilde{d}_{t}=-1,\qquad\quad\,\,\,\,\,\, \,\tilde{d}_{d}=1,\nonumber\\
&&\bar{\lambda}_{i}=\left(R_{i1}\, s_{\beta}+R_{i2}c_{\beta}\right)\,\text{Re}\,a_{2\Sigma}-R_{i3}\,\text{Im}\,a_{2\Sigma}\,.\nonumber
\end{eqnarray}
In equations above the matrix $R$ is defined to diagonalize the neutral bosons mass matrix $R M_{\text{neutral}}^2 R^T~=~\diag(m_{h_1}^2,m_{h_2}^2,m_{h_3}^2,m_{h_4}^2)$. In terms of matrix $R$ the weak eigenstates are related to mass eigenstates via $(H_1^0,H_2^0,A^0,\Sigma^0)=(h_1,h_2,h_3,h_4)\mcdot R$\,. Note that the fermions directly couple to the neutral Goldstone boson $G^0$, while the scalar interaction $\Sigma^+\Sigma^- G^0$ is absent at tree level. 
\subsection{Phenomenological parameters}
\begin{table}
\caption{\label{tab:table1} Table of the parameters in the potential versus the phenomenological parameters.  }
\begin{ruledtabular}
\begin{tabular}{lcr}
\scriptsize{Parameters in the potential}&\scriptsize{Phenomenological parameters} \\
\hline
$\lambda_1, \lambda_2, \lambda_3,\lambda_4, \text{Re}\lambda_5$ & $v,\tan\beta, \alpha,\text{Re}\,a_{2\Sigma}, \text{Re}\,a_{2S}$\\
$ m_{11}^2, m_{22}^2\,,\text{Re}a_{2\Sigma}, \text{Im} a_{2\Sigma}, \mu_{\Sigma}$ &$\delta_{\Sigma}, \delta_{S}, b_{4\Sigma}, b_{4S},m_{\Sigma}, m_{S}$\\
 $ b_{4\Sigma},\text{Re}a_{2S}, \text{Im} a_{2S}, b_{2S},b_{4S}$ & $m_{H^+}, m_{h_1},m_{h_2},m_{h_3}$ \\
\end{tabular}
\end{ruledtabular}
\end{table}

We are interested in the zero-temperature relations among the physical parameters of the theory such as masses of neutral and charged scalars, and the parameters in the potential. In Table~\ref{tab:table1} we list the set of parameters of our potential and the phenomenological parameters. 

In the 2HDM sector the CPV is absent because we take $m_{12}^2=0$, and therefore $\text{Im}\lambda_{5}=0$ from  the minimization conditions in \eq{eq:minimizationconditions}. The only two CPV phases in our theory are due to the triplet and singlet and are represented by $\delta_{\Sigma}, \delta_S$.

Motivated by present fits on to Higgs observables (see, {\em e.g.}, \cite{Inoue:2014nva} ), we assume the SM alignment limit $\alpha=\beta-\pi/2$, in which the couplings $\lambda_1,\dots \lambda_5$ are related to phenomenological parameters via \footnote{These expressions are identical to analogous ones for the pure 2HDM case \cite{Inoue:2014nva} with additional assumptions $m_{12}^2=\text{Im}\lambda_5=\alpha_b=\alpha_c=0, \alpha=\beta-\pi/2$.}
\begin{eqnarray}
&&\lambda_1=\frac{m_{h_1}^2+m_{h_2}^2\tan^2\beta}{v^2}\,,\nonumber\\
&&\lambda_2=\frac{m_{h_1}^2+m_{h_2}^2\cot^2\beta}{v^2}\,,\nonumber\\
&&\lambda_3=\frac{m_{h_{1}}^2-m_{h_2}^2+2m_{H^+}^2}{v^2}\,,\nonumber\\
&&\lambda_{4}=\frac{m_{h_3}^2-2m_{H^+}^2}{v^2}\,,\nonumber\\
&&\text{Re}\,\lambda_5=-\frac{m_{h_3}^2}{v^2}\,,\nonumber\\
&&m_{11}^2=m_{22}^2=m_{h_1}^2\,.
\end{eqnarray}

The matrix $R_{ij}$ that enters the Feynman rules in \eq{eq:feynmanrulescouplings} in general is a function of three angles  $\alpha, \alpha_b, \alpha_c$
\begin{eqnarray}
&&R(\alpha_c,\alpha_b,\alpha)=R_{23}(\alpha_c)R_{13}(\alpha_b)R_{12}\left(\alpha+\frac{\pi}{2}\right).\nonumber
\end{eqnarray}
In the case of interest for us we have
\begin{eqnarray}
R(0,0,\beta-\pi/2)=\left( \begin{array}{cccc}
c_{\beta} &s_{\beta}&0  & 0\\
-s_{\beta}& c_{\beta} &  0 & 0 \\
0 & 0 & 1&0\\
0  & 0 & 0 &1 \end{array} \right)\,.
\end{eqnarray}
Finally $\mu_{\Sigma}^2$ is found from \eq{eq:chargedmasses}, $b_{2S}$ is found from \eq{eq:singletmassformula}, and the imaginary parts of $a_{2\Sigma}, a_{2S}$  are found from their real parts and the angles $\delta_\Sigma, \delta_{S}$\,.

The formulae described in this subsection allow us to recover all the parameters in the potential starting from phenomenological parameters in Table~\ref{tab:table1} . 

\section{Observables}
\label{sec:pheno}
In this section we review two observables that place constraints on our illustrative scenario: EDMs and the Higgs diphoton decay rate. The former constrains the interplay of masses and the CP-violating phase $\delta_\Sigma$. The latter is sensitive to the magnitude of Higgs portal coupling $a_{2\Sigma}$ and the triplet masses.  

\subsection{Electric Dipole Moments}
EDMs of non-degenerate systems are CP-violating observables, and EDM searches provide constraints on BSM sources of CPV. Treating the SM as an effective field theory, new sources of CPV can be characterized by dimension-6 operators, which include elementary fermion EDMs, quark chromo-EDMs, the CPV 3-gluon operator, etc. (see review \cite{Engel:2013lsa}). In general, a model-independent analysis of EDM limits would require us to consider a large set of these CPV operators. Here, we can focus on the electron and quark EDM operators, as these are the only experimentally relevant dimension-6 operators that can be generated in our model at lowest non-trivial  (2-loop) order\footnote{In principle, there also exist one loop EDMs generated by scalar exchange; however, these contributes are highly suppressed by the light fermion Yukawa couplings}.

\begin{figure}
\includegraphics[scale=0.5]{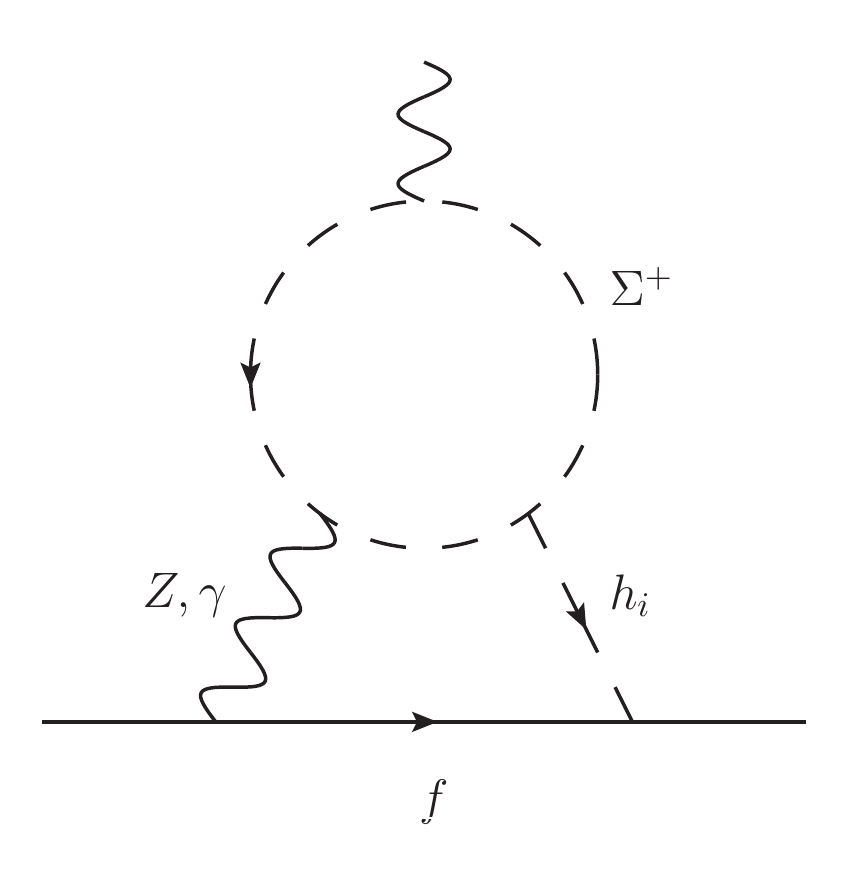}
\caption{\label{fig:BarrZee} The dominant EDM contribution comes from these Barr-Zee diagrams and their mirror graphs.}
\end{figure}

We have assumed that our 2HDM sector is CP-conserving, and that the new sources of CPV involve either the triplet or the singlet. As a result, the leading contribution to SM fermion EDMs comes from graphs involving the charged scalar $\Sigma^{+}$ shown in Figure~\ref{fig:BarrZee}.
These so-called Barr-Zee diagrams are familiar from EDM analyses in 2HDM, and the result for dimensionless EDM $\delta_f\equiv-v^2d_f/2m_fe$ (as in \cite{Engel:2013lsa}) from this diagram is analogous to the $H^+$ loop result \cite{Kao:1992jv}:
\begin{eqnarray}
&&\left(\delta_f\right)_{\Sigma^+}^{h\gamma\gamma}=\frac{Q_fe^2 v^2}{256\pi^4 m^2_{\Sigma^+}}\sum_{i=1}^{3}\left[f(z_\Sigma^i)-g(z_\Sigma^i)\right]\bar{\lambda}_i\tilde{c}_{f,i}\,,\nonumber\\
&&\left(\delta_f\right)_{\Sigma^+}^{hZ\gamma}=\frac{g^V_{Z\bar{f}f}g_2\,c_W v^2}{256\pi^4 m^2_{\Sigma^+}}\\
&&\times\sum_{i=1}^{3}\left[\tilde{f}(z^i_{\Sigma},m^2_{\Sigma^+}/M_Z^2)-\tilde{g}(z_\Sigma^i,m^2_{\Sigma^+}/M_Z^2)\right]\bar{\lambda}_i\tilde{c}_{f,i}\,,\nonumber
\end{eqnarray}
where the vector coupling of the electron to the $Z$ equals $g^V_{Z\bar{e}e}~=~g_2(s_W^2 -~1/4)/c_{W}$ and
\begin{eqnarray}
z_\Sigma^i\equiv  m_{\Sigma^+}^2/m_{h_i}^2.
\end{eqnarray}
The loop functions $f(z), g(z), \tilde f(z), \tilde g(z)$ are listed in Appendix~A.
Analyzing expressions for $\bar{\lambda}_i$ in \eq{eq:feynmanrulescouplings} in the limit $\alpha_b=\alpha_c=m_{12}^2=\text{Im}\,\lambda_5=0$, we see that only $i=3$ contributes to the fermion EDM. Thus theoretical predictions for both electron and neutron EDMs depend only on the masses $m_{h_3}, m_{\Sigma^+}$\,. The electron EDM which gives the strongest bound is proportional to the $\sin\delta_{\Sigma}$ and $\tan\beta$ parameters: $\delta_e\sim\sin\delta_{\Sigma}\tan\beta$.

Up and down quark EDMs contribute to the neutron/proton EDM as
\begin{equation}
d_{n/p} = e \sum_{q = u,d} \zeta_{n/p}^{q} \delta_{q},\label{eq:edm}
\end{equation}
where $\zeta_{n/p}^{q}$ are matrix elements of quark EDM operators between neutron/proton wavefunctions. They equal
\begin{eqnarray}
 \zeta_{n/p}^{q}=-\frac{2m_q\rho_{n/p}^{q}}{v^2},\label{eq:edmmatrixelements}
\end{eqnarray}
where the dimensionless tensor charges $\rho_{p/n}^q$ have been evaluated in Ref.\cite{Bhattacharya:2015wna} and we use them below in section \ref{subsec:ResultsEDMs}.


While the down EDM is proportional to $\tan\beta$ like the electron EDM, the up EDM behaves as $\cot\beta$ due to the different Yukawa structure. This means that the neutron and proton EDMs can potentially place a stronger bound on the low $\tan\beta$ region in comparison with the electron EDM bound.


\subsection{Higgs Diphoton Decay}
CMS \cite{Khachatryan:2014ira} and ATLAS \cite{Aad:2014eha} measurements of the Higgs diphoton decay are consistent with SM and provide bounds $a_{2\Sigma}$ and $m_\Sigma$. The observed signal strength for the Higgs diphoton decay channel, divided by the corresponding SM value, is defined as $\mu$. Approximating the SM value of the diphoton decay signal strength with contributions from the top and $W-$boson loops, using standard results from Ref.\cite{Gunion:1989we} we obtain the following theoretical prediction for the value of $\mu$
\begin{eqnarray}
\mu^{\text{theory}}_{2\text{HDM}\Sigma}=\frac{\left |3\, g_t\,Q_t^2 F_{1/2}(\tau_t)+g_W\,F_1(\tau_{W})+g_{\Sigma}\,F_0(\tau_{\Sigma})\right |^2}{\left |3\,Q_t^2 F_{1/2}(\tau_t)+F_1(\tau_{W})\right |^2},\nonumber\\ \label{eq:analysis1}
\end{eqnarray}
where $\tau_t=4m_t^2/m_{h_1}^2, \tau_W=4m_W^2/m_{h_1}^2, \tau_{\Sigma}=4m_{\Sigma}^2/m_{h_1}^2$ and \cite{Gunion:1989we}
\begin{eqnarray}
&&F_{1/2}(\tau)=-2\tau[1+(1-\tau)f(\tau)],\nonumber\\
&&F_1(\tau)=2+3\tau[1+(2-\tau)f(\tau)],\nonumber\\
&&F_0(\tau)=\tau[1-\tau f(\tau)]\,.
\end{eqnarray}
Function $f(\tau)$ reads \cite{Gunion:1989we}
\begin{eqnarray}
f(\tau)=\Bigg\{ \begin{array}{cc}
\arcsin^2(1/\sqrt{\tau}), \,\qquad\,\,\,\,\,\,\,\,\,\,\,\text{if } \tau\ge1  \\
-\frac{1}{4}\left(\ln\frac{1+\sqrt{1-\tau}}{1-\sqrt{1-\tau}}-i\pi\right)^2,\,\,\text{if }\tau<1    \,.\end{array}
\end{eqnarray}
The coefficients $g_t, g_W, g_{\Sigma}$ are
\begin{eqnarray}
&&g_t=1,\label{eq:analysis2}\\
&&g_W=1,\nonumber\\
&&g_{\Sigma}=\frac{1}{2}\sin2\beta\,\text{Re}\,a_{2\Sigma}\left(\frac{v}{m_{\Sigma}}\right)^2\,.\nonumber
\end{eqnarray}
The numerical constraints from the Higgs to diphoton data on the parameters $\tan\beta, m_{\Sigma}$ are presented in the section~\ref{sec:Higgsdiphoton} below.

\section{Baryogenesis and two-step phase transition}
\label{sec:transport}
We now analyze the dynamics of BAU generation during the first step of the 2SPT process. Following early work, we first compute the net number density of LH fermions $n_L$ generated by CP-violating interactions at the bubble walls (the space-time varying scalar field vevs). As it diffuses ahead of the bubble wall, the non-vanishing $n_L$ then catalyzes creation of a non-vanishing baryon number density by the EW sphalerons. Since the EW sphaleron rate $\Gamma_\mathrm{ws}$ is typically much slower than the rates for processes that govern $n_L$ generation, we treat the computations of $\rho_B$ and $n_L$ as separate steps. The latter entails deriving and solving a coupled set of transport equations of the form:
\be
\partial_\lambda j^\lambda_k = -\sum_A \, \Gamma_A\left(\mu_k - \mu_\ell -\cdots\right) + S_k^{\text{\CPV}}\ \ \ ,
\ee
where $j_k^\lambda$ and $\mu_k$ are the number density current and chemical potential for particle species \lq\lq $k$", $\Gamma_A$ are a set of particle number changing reaction rates that involve species $k$ and other species relevant to the problem, and $S_k^{\text{\CPV}}$ is a CP-violating source for species $k$. 

We derive these equations using the Schwinger-Keldysh closed time path (CTP) framework that is appropriate for out-of-equilibrium dynamics at finite temperature \cite{Schwinger:1960qe,Mahanthappa:1962ex,Bakshi:1962dv,Bakshi:1963bn,Keldysh:1964ud,Chou:1984es}.
For a detailed review of the formalism in the context of the EWBG in SUSY see {\em e.g.} Refs.~\cite{Lee:2004we,Cirigliano:2006wh}\,. For brevity we provide the main results here without going into the details of the formalism. As this work provides an initial study of the viability of two-step EWBG, we employ several approximations and assumptions to make the computation reasonably tractable, deferring a more exhaustive treatment to future work:
\begin{itemize}
\item We take the bubble walls to be planar, and treat all quantities only as functions of the co-moving co-ordinate $z=x+v_w t$, where $v_w$ is the wall velocity and $x$ is the position relative to the center of the wall. We illustrate the dependence of $Y_B$ on $v_w$, and give illustrative results for a value of $v_w$ that is within the range of values obtained from EWBG studies for other models. 
\item We use bubble wall profiles, $v_\Sigma(x)$ and $v_S(x)$, that have the typical form obtained in other studies of bubble walls, again showing the dependence of $Y_B$ on the parameters that characterize the profile and choosing typical values for purposes of illustration.
\item We compute the CPV sources using the vev-insertion approximation (VIA), which amounts to expanding the mass-squared matrix to second order in the off-diagonal elements $m_{12}^2(x)$. The sources then depend on the interference of these elements at two different space-time points, {\em viz}, $m_{12}^2(x) m_{12}^2(y)^\ast$. As we will see below, when the CPV phase in the $m_{12}^2(x)$ vary with space-time, the interference $m_{12}^2(x) m_{12}^2(y)^\ast$ contains a non-vanishing CPV phase. 

\end{itemize}

The VIA assumes that the particle-antiparticle asymmetry generation is dominated by the region near the phase boundary, where the vevs are small compared to both $T$ and the difference $|m_{11}^2-m_{22}^2|^{1/2}$. While the approximation is thought to provide a reasonable estimate of the magnitude of the CPV sources, it is associated with theoretical uncertainties. In particular, it neglects the impact of flavor oscillations, which become important in the region where the off-diagonal term in $\mathbf{M}^2 $ is comparable to $|m_{11}^2-m_{22}^2|$. In the present instance, the potential $V(H_1,H_2)$ contains no tree-level contribution to $m_{12}^2$ that would give rise to flavor oscillations. Flavor oscillations will, nevertheless, be induced by a finite-temperature contribution as well as the non-vanishing $v_\Sigma$ and $v_S$. Thus, we will treat our results for parameter choices in this region with a healthy dose of salt.

A more complete treatment that includes flavor oscillations requires a resummation of the vevs to all orders, as well as a  first principles treatment of the thermalizing interactions between the scalar fields and the other particles in the finite-T plasma. Initial efforts to carry out the former have been performed in Refs.~\cite{Carena:2000id,Carena:2002ss,Konstandin:2004gy, Konstandin:2005cd}, focusing on the $Y_B$ in the minimal supersymmetric Standard Model as generated by gaugino-Higgsino interactions with the bubble walls (for an extensive discussion and related references, see Ref.~\cite{Morrissey:2012db}). The authors of Refs.~\cite{Konstandin:2004gy, Konstandin:2005cd} found a significant reduction in the asymmetry compared to the result in the vev-insertion approximation. However, as later pointed out in Ref.~\cite{Cirigliano:2011di}, the computation of Refs.~\cite{Konstandin:2004gy, Konstandin:2005cd} neglected the effects of diffusion ahead of the bubble wall and dropped the dominant CPV source for the flavor-diagonal particle number densities. Retaining the latter and including the effects of diffusion leads to a \lq\lq resonant-enhancement" of $Y_B$ in the small $|m_{11}^2-m_{22}^2|$ regime that is consistent with what is observed in the VIA. 
Consequently, we conclude that for our present purpose of evaluating the viability of two-step EWBG the vev-insertion approximation provides a reasonable estimate of the magnitude of the baryon asymmetry one might anticipate in this scenario. We defer to future work an analysis using the more numerically intensive framework of Refs.~\cite{Cirigliano:2009yt,Cirigliano:2011di} that also requires modeling in detail the CP-conserving thermalizing interactions in the high-temperature plasma.

\subsection{Relevant interactions}
With these comments in mind, we now provide the Lagrangian for interactions that are responsible for the BAU generation below. All scalar interactions are contained in the effective potential $V(H_1,H_2,\Sigma, S)$, see equations \eq{eq:deltaV},\,\eq{eq:twohiggsdoubletlagrangian}\,.  The source terms are generated by the interactions that have two Higgs fields $H_1, H_2$ and two VEVs of the triplet and singlet fields $v_{\Sigma}^2, v_{S}^2$. The corresponding Lagrangian is given by
\begin{eqnarray}
\mathcal{L}_{\text{scalar}}^{S}=-\frac{\left(a_{2\Sigma}\,v_{\Sigma}^2+a_{2S}\,v_{S}^2\right)}{2}\,\left[H_{1}^{\dagger}H_{2}+\text{h.c.}\right]\,.
\end{eqnarray}

The CP-conserving interactions include both scattering and particle number changing reactions. The former determine how effectively particle asymmetries generated at the bubble wall diffuse into the broken phase where the weak sphalerons are unsuppressed. We take these into account using the diffusion {\em ansatz} (see below). The leading particle number changing reactions are mediated by tri-scalar and Yukawa interactions. For the former the following interaction Lagrangian is needed 
\begin{eqnarray}
\mathcal{L}_{\text{scalar}}^{Y}=-\left(a_{2\Sigma}\,v_{\Sigma}\,\Sigma^0+a_{2S}\,v_{S}\,S\right)\left[H_{1}^{\dagger}H_2+\text{h.c.}\right]\,.\nonumber\\
\end{eqnarray}
For the latter, the Yukawa Lagrangian in the type-II 2HDM is given in \eq{eq:YukawaLagrangian}\,. Keeping only terms proportional to the top Yukawa coupling we arrive at
\begin{eqnarray}
\mathcal{L}_{\text{fermion}}^Y=y_t\left(-H_2^0\,\bar{t}_Rt_L+H_2^+\,\bar{t}_Rb_L\right)\,,\label{eq:LYFermion}
\end{eqnarray}
where $y_t$ is related to the $(3,3)$ component of the Yukawa matrix $Y_U$ and the top quark mass in the following way $y_t\equiv Y_U^{33}=\sqrt{2}m_t/|v_2|$\,. The corresponding expression for the bottom Yukawa coupling is $y_b\equiv Y_D^{33}=\sqrt{2}m_b/|v_1|$\,. Note that in the regime of moderate to large $\tan\beta$, wherein $y_b$ is enhanced over its SM value, explicit inclusion of bottom Yukawa interactions can be decisive\cite{Chung:2008aya}.

\subsection{The source term}
\begin{figure}[!t]
\includegraphics[width=4cm]{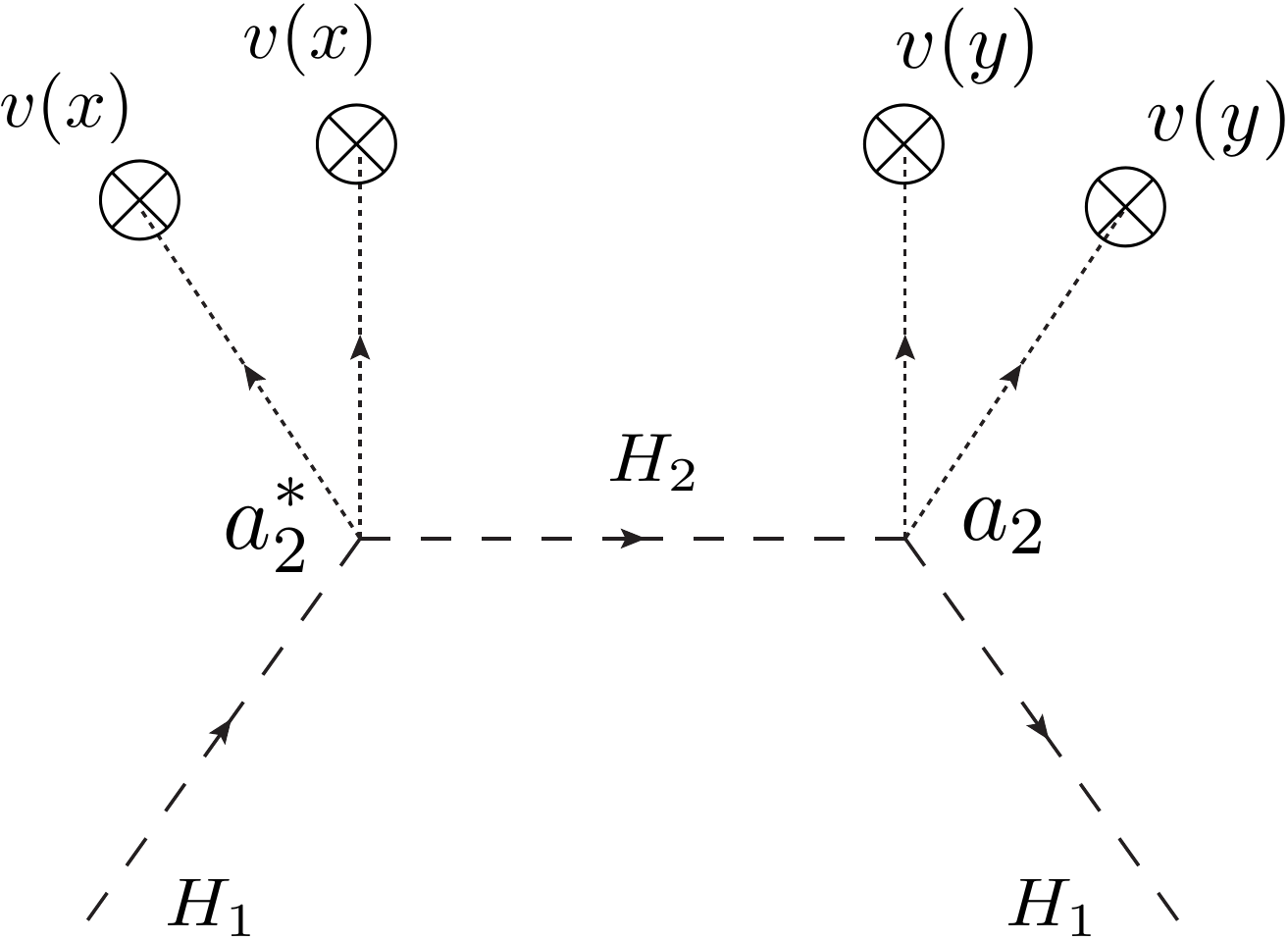} \quad\includegraphics[width=4cm]{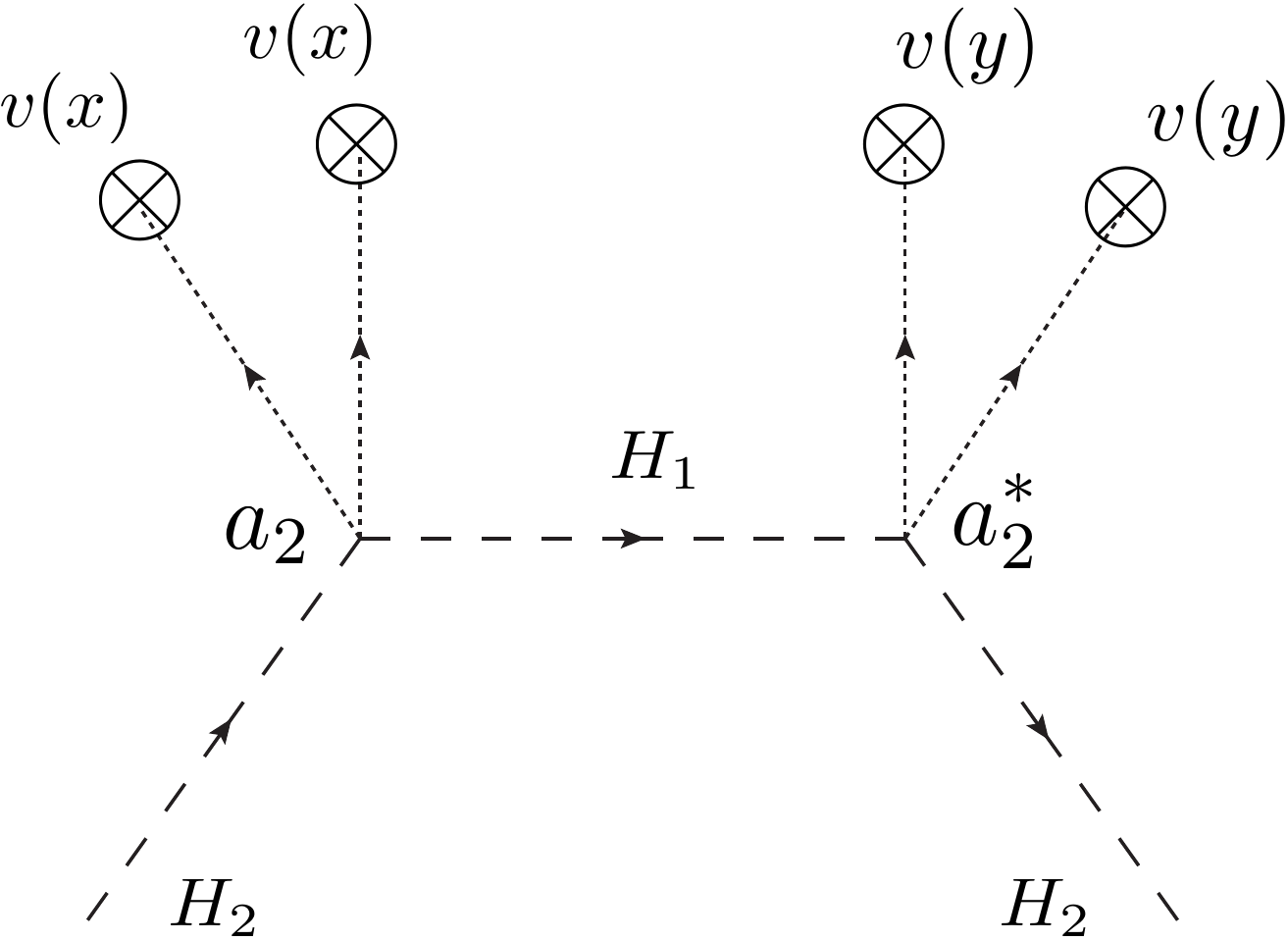}
\caption{VEV insertion Feynman graphs in the 2HDM with an addition of a triplet and a singlet. The VEV and the $a_2$ coupling in the graphs correspond to the triplet and a singlet correspondingly.}\label{fig:realVEV}
\end{figure}

Under the VIA, the processes shown in Figure~\ref{fig:realVEV} generate the source term densities $H_1, H_2$. The complex couplings that enter the computation are $a_{2\Sigma}, a_{2S}$. As a result the source term has two qualitatively different parts: the CP conserving and the CP violating. Applying straightforwardly the CTP method\cite{Lee:2004we}, we obtain a CP conserving term 
\begin{eqnarray}
&&S^{\text{CP}}_{H_1}=\Gamma_{M}^{+}\left(\mu_{H_1}+\mu_{H_2}\right)+\Gamma_{M}^-\left(\mu_{H_2}-\mu_{H_1}\right),\nonumber\\
&&S^{\text{CP}}_{H_2}=-S^{\text{CP}}_{H_1}\,,
\end{eqnarray}
where the CP conserving relaxation rates $\Gamma_{M^\pm}$ are
\begin{eqnarray}
\Gamma_{M^\pm}&=&-\frac{3\,W^\pm}{2\pi^2 T^3}\left|a_{2\Sigma} v_{\Sigma}^2(x)+a_{2S} v_{S}^2(x)\right|^2\,.\nonumber\\ \label{eq:GammaM}
\end{eqnarray}
The role of the CP conserving term in the Boltzmann equations is that of the relaxation type. The CP violating source term equals
\begin{eqnarray}
S_{H}^{\text{\CPV}}(x)&=&\frac{|a_{2\Sigma}a_{2S}|\sin(\delta_S-\delta_{\Sigma})}{\pi^2}v_S(x)v_{\Sigma}(x)\nonumber\\
&&\times \left[v_S(x)\dot{v}_{\Sigma}(x)-\dot{v}_S(x)v_{\Sigma}(x)\right]\Lambda\,.\label{eq:sourcetermformula}
\end{eqnarray}
The role of the CPV source term in the Boltzmann equations is to generate the particle-antiparticle asymmetries. The thermal functions $W^{\pm}, \Lambda$ depend on the temperature $T$ and thermal masses of Higgs bosons $H_{1}, H_{2}$ and are given by 
\begin{small}
\begin{eqnarray}
&&W^\pm=\int\frac{k^2dk}{\omega_1\omega_2}\frac{1}{2}\text{Im}\left(\frac{h_B(\epsilon_2)\mp h_B(\epsilon_1^*)}{\epsilon_2-\epsilon_1^*}-\frac{h_B(\epsilon_2)\mp h_B(\epsilon_1)}{\epsilon_2+\epsilon_1}\right),\nonumber\\
&&\Lambda=\int\frac{k^2d k}{\omega_1\omega_2}\text{Im}\left(\frac{n_B(\epsilon_1^*)-n_B(\epsilon_2)}{(\epsilon_1^*-\epsilon_2)^2}+\frac{n_B(\epsilon_1)+n_B(\epsilon_2)+1}{(\epsilon_2+\epsilon_1)^2}\right)\,,\nonumber\\ \label{eq:Wlambdaformulas}
\end{eqnarray} 
\end{small}
where $\epsilon_i=\sqrt{k^2+m_{H_i}^2}-i\Gamma_{H_i},\, i=1,2$\,. The functions $W^\pm, \Lambda$ have a well known resonant enhancement when $m_{H_1}=m_{H_2}$ \cite{Lee:2004we}, which we stress below in the Section~\ref{eq:resultsbaryo}. 

\subsection{Particle Number Changing Rates}
\begin{figure}[!t]
\includegraphics[width=4cm]{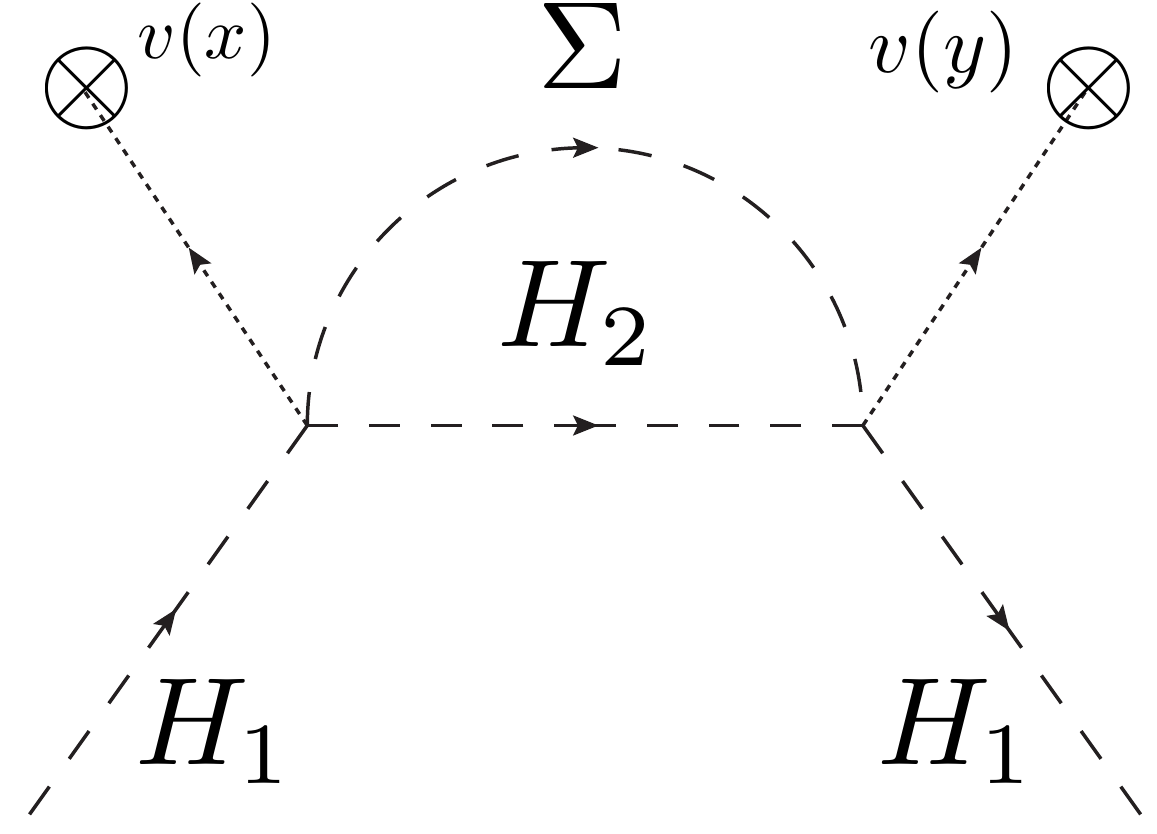}\quad \includegraphics[width=4cm]{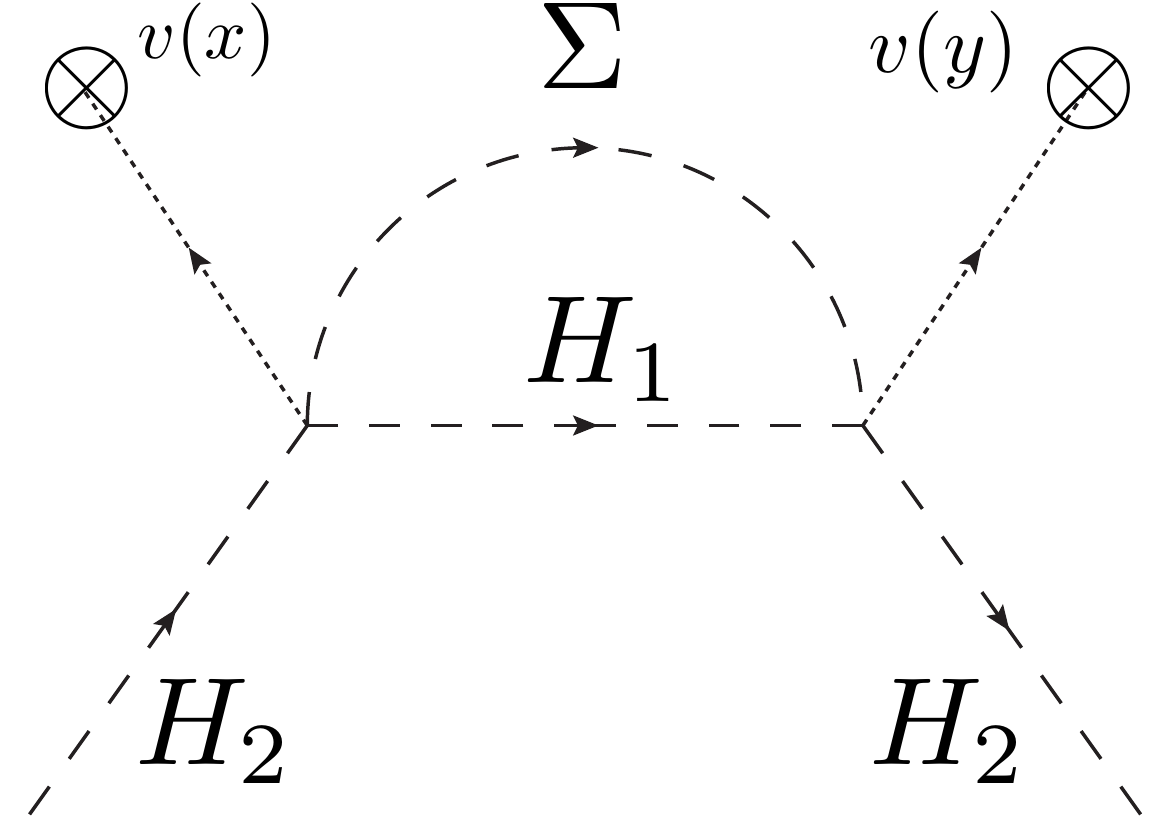}\\
\includegraphics[width=4cm]{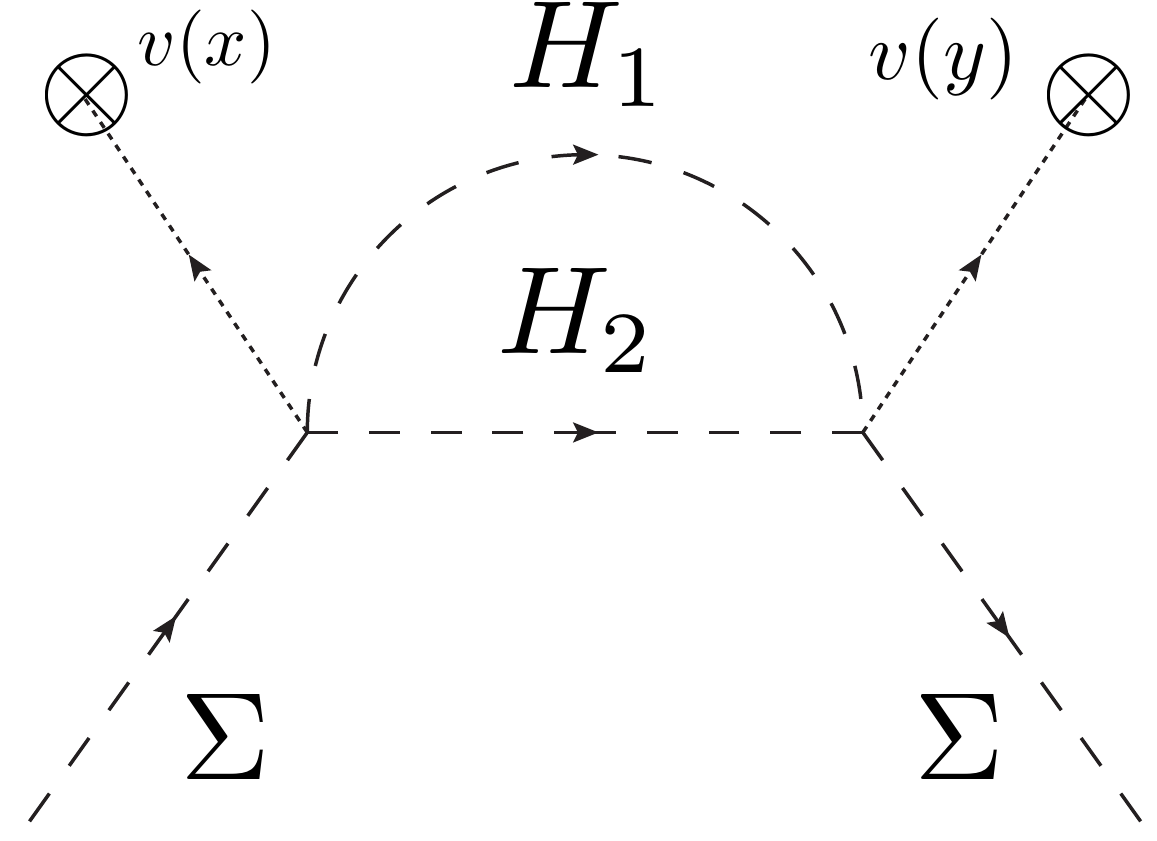}
\caption{First two graphs define the relaxation rate $\Gamma_H$. The last graph leads to a zero source term for the $\Sigma^0$ field.}\label{fig:realSelfEnergy}
\end{figure}
The first two graphs shown in Fig.~\ref{fig:realSelfEnergy} evaluated using the CTP formalism lead to terms in the transport equations for the currents of the densities $H_1, H_2$: $\partial_{\mu}j_{H_i}^{\mu}=~\mp~(\mu_{H_1}-~\mu_{H_2})\Gamma_H$ with $i=1, 2$, where
\begin{eqnarray}
&&\Gamma_H\approx\frac{12}{T^2}\,\Big[|a_{2\Sigma}|^2\mathcal{I}_{B}\left(v_{\Sigma}(x);m_{H_1},m_{H_2},m_{\Sigma}\right)\nonumber\\
&&\qquad\qquad+|a_{2S}|^2\mathcal{I}_{B}\left(v_S(x);m_{H_1},m_{H_2},m_{S}\right)\Big]\,.
\end{eqnarray}
The approximate sign indicates the assumption of slowly varying  triplet and the singlet vevs in the first two graphs of Figure~\ref{fig:realSelfEnergy}: $v_{\Sigma, S}(x)\approx v_{\Sigma,S}(y)$\,. The function $\mathcal{I}_{B}$ depends on the temperature and thermal masses of $H_1, H_2, \Sigma$ and $S$ fields, it can be found in Ref. \cite{Cirigliano:2006wh} and is not repeated here for brevity. The third graph in Figure~\ref{fig:realSelfEnergy} equals to zero in the CTP formalism, which is manifestation of the fact that the $\Sigma^0$ field has a zero chemical potential.
\begin{figure}[!t]
\includegraphics[width=4cm]{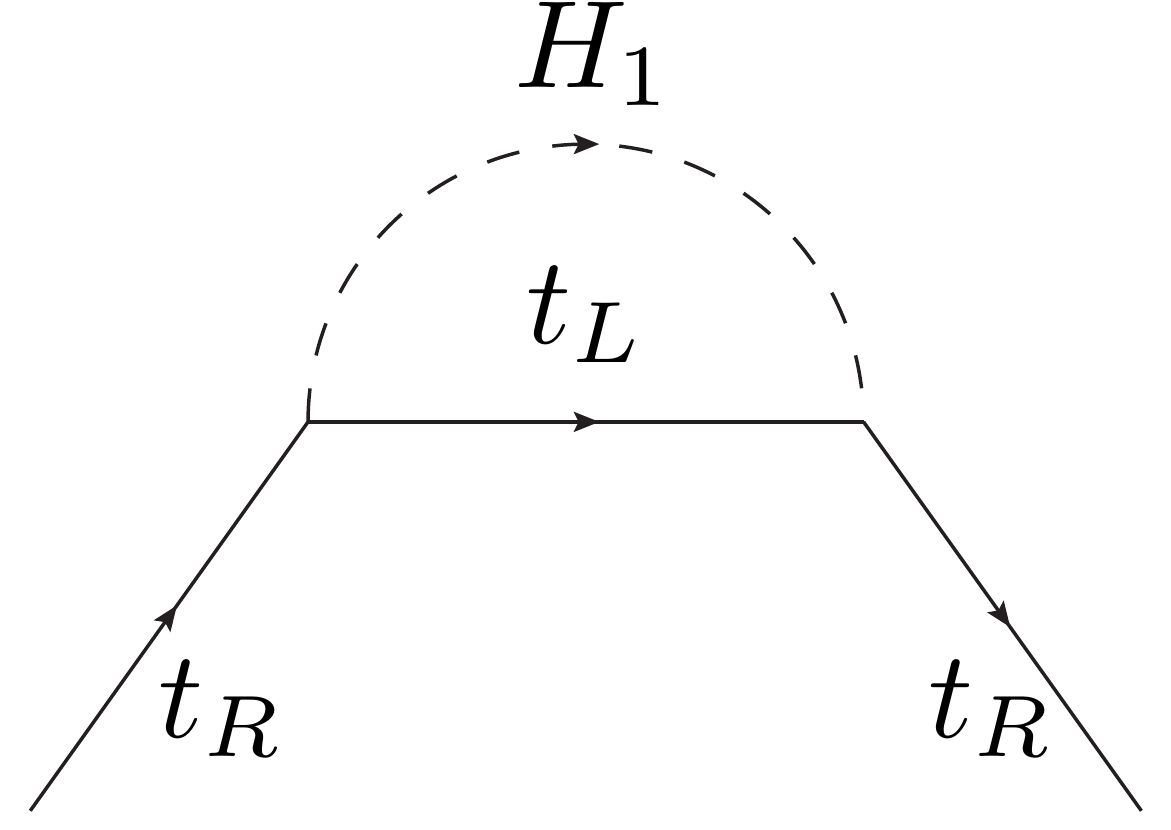} \quad\includegraphics[width=4cm]{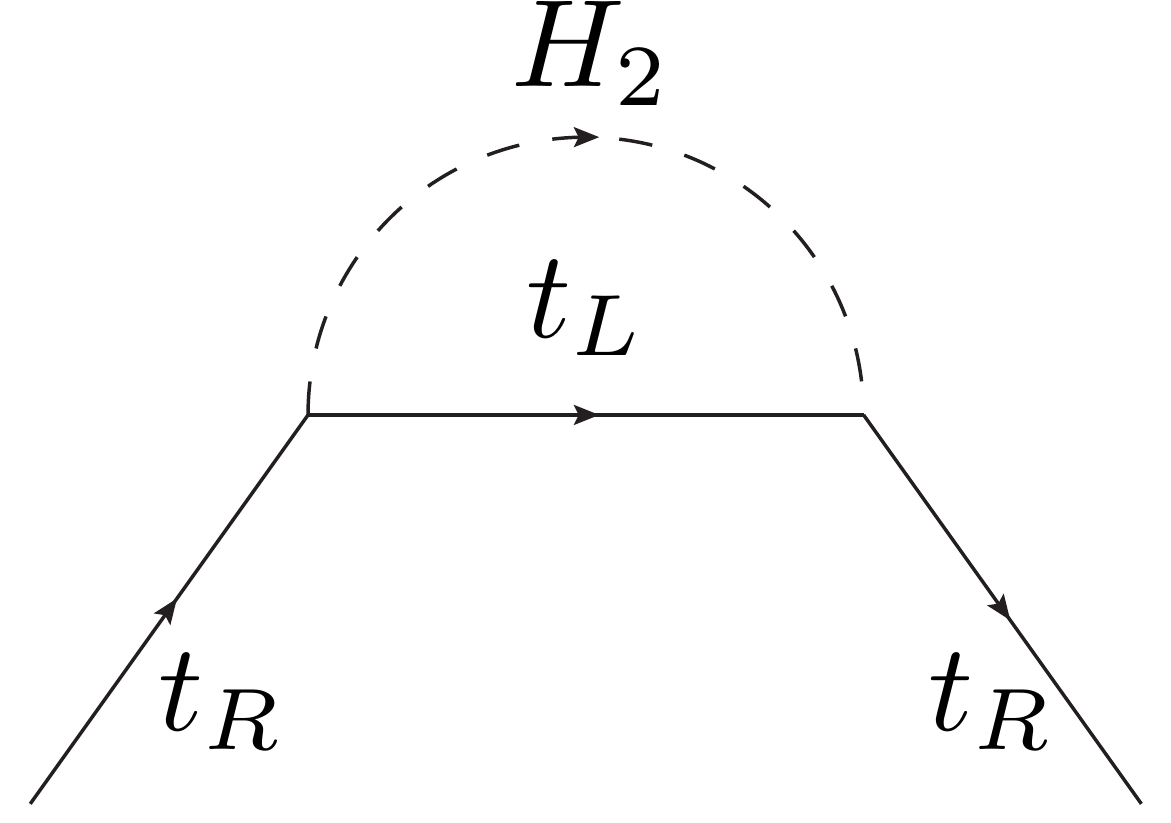}\\
\includegraphics[width=4cm]{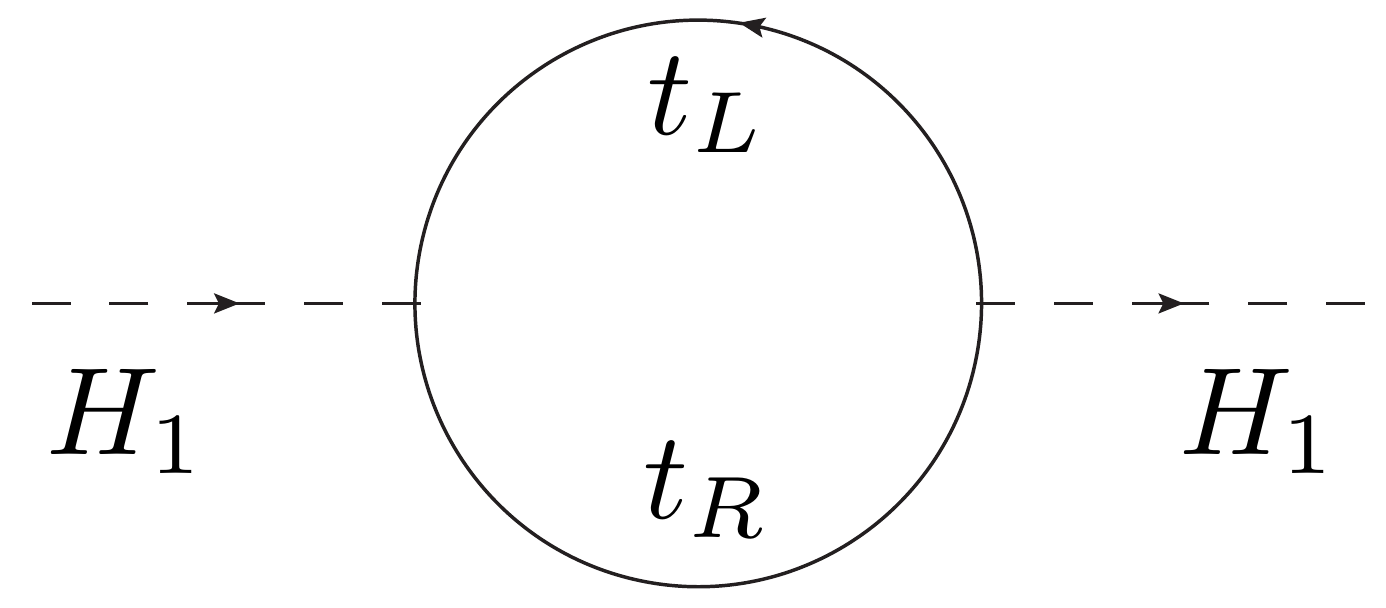}\quad\includegraphics[width=4cm]{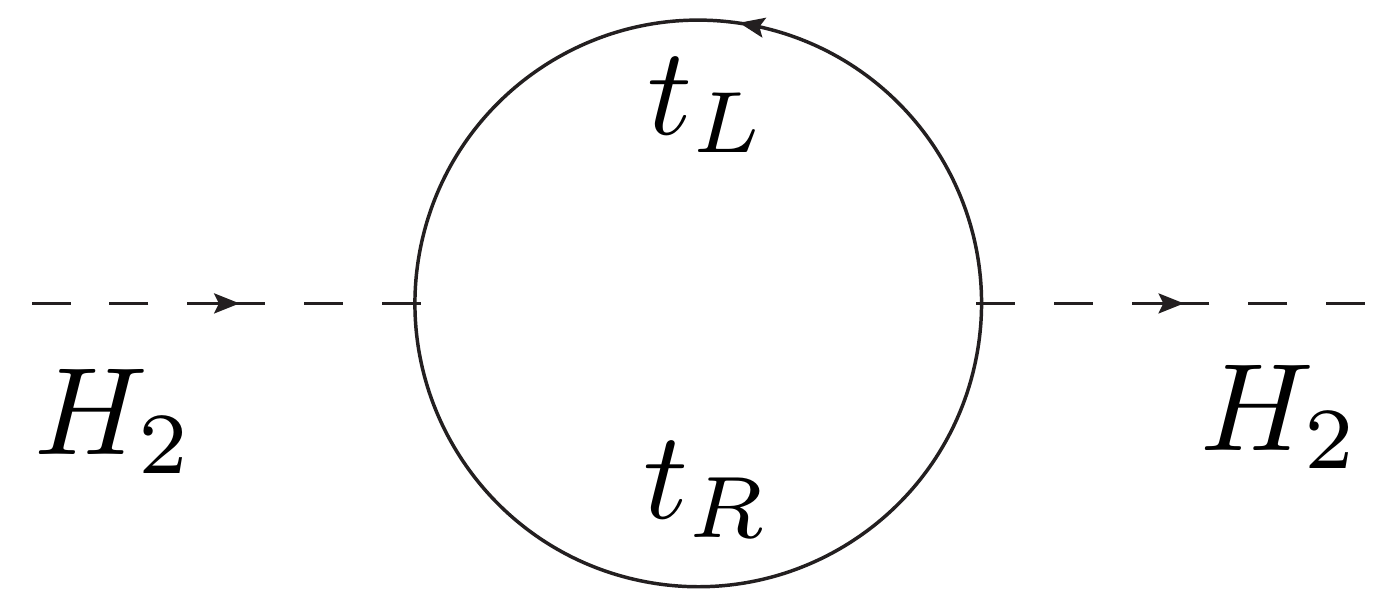}
\caption{Processes in the plasma induced by the Yukawa terms. }\label{fig:realYukawa}
\end{figure}

In Figure \ref{fig:realYukawa} we show the additional interactions in the plasma that lead to Yukawa $\Gamma_Y$ and corresponding additional terms in the transport equations that arise from Yukawa interactions. These processes play important role  as they transfer the particle-antiparticle asymmetry generated in the scalar sector to the left handed quark anti-quark asymmetry. The type~II 2HDM has a Yukawa sector equivalent to MSSM, if one excludes the superparticles from the latter Lagrangian. Thus the relaxation rate $\Gamma_Y$ for our scenario and the corresponding terms in the transport equations can be readily obtained from MSSM calculations \cite{Cirigliano:2006wh}:
\begin{eqnarray}
&&\Gamma_{Y}=\frac{12 N_C\,y_t^2}{T^2}\,\mathcal{I}_F\left(m_{t_R},m_Q,m_{H_2}\right)+0.129\frac{g_3^2}{4\pi}T\,,\nonumber\\ \label{eq:GammaY}
\end{eqnarray}
where the function $\mathcal{I}_F$ can be found in Ref.~\cite{Cirigliano:2006wh} and will not be repeated here. The additional term $0.129\frac{g_3^2}{4\pi}T$ is included \cite{Joyce:1994zn,Chung:2009qs} as an estimate of the four-body contributions to the $\Gamma_Y$, which plays role  in some in the parameter space regions where, due to kinematic threshold effects, $\mathcal{I}_F$ vanishes. In particular,  $\mathcal{I}_F$ vanishes for $m_{H_2}<150\,\text{GeV}$, and for $m_{H_2}=200\,\text{GeV}$, the first term in the \eq{eq:GammaY} $\approx 3\text{GeV}$. This to be compared to the value of the four-body term $\approx 1.9\,\text{GeV}$\,.

We use the standard form for the strong sphaleron rate~\cite{Huet:1995sh,Giudice:1993bb,Moore:1997im}
\begin{eqnarray}
\Gamma_{ss}=16\,\kappa\,\alpha_s^4\,T,
\end{eqnarray}
with $\kappa\simeq 1$ and $\alpha_s$ is the strong coupling.

\subsection{Boltzmann equations}
In order to analyze all the transport processes that arise from CTP processes considered in the above subsection, we construct the following four densities:
\begin{eqnarray}
&& T=n_{t_R},\nonumber\\
 &&Q=n_{t_L}+n_{b_L},\nonumber\\
 &&H=n_{H_{2}^+}+n_{H_{2}^0}-n_{H_{1}^+}-n_{H_{1}^0},\nonumber\\
 &&h=n_{H_{1}^+}+n_{H_{1}^0}+n_{H_{2}^+}+n_{H_{2}^0}\,,
\end{eqnarray}
where $T,Q,H,h$ correspond to the densities of right handed quarks, left handed quarks and two linear combinations of $H_1, H_2$ densities, respectively. The resulting set of the transport equations are
\begin{small}
\begin{eqnarray}
&&\partial^{\mu}T_{\mu}=-\Gamma_Y\left[\frac{T}{k_T}-\frac{Q}{k_Q}-\left(\frac{H}{k_H}+\frac{h}{k_h}\right)\right]\nonumber\\
&&\qquad\qquad+\Gamma_{ss}\left(\frac{2Q}{k_Q}-\frac{T}{k_T}+\frac{9(Q+T)}{k_B}\right)\,,\nonumber\\
&&\partial^{\mu}Q_{\mu}=-\Gamma_Y\left[\frac{Q}{k_Q}-\frac{T}{k_T}+\left(\frac{H}{k_H}+\frac{h}{k_h}\right)\right]\nonumber\\
&&\qquad\qquad-2\Gamma_{ss}\left(\frac{2Q}{k_Q}-\frac{T}{k_T}+\frac{9(Q+T)}{k_B}\right)\,,\nonumber\\
&&\partial^{\mu}H_{\mu}=-\Gamma_{M}^+\frac{h}{k_h}-\left(\Gamma_M^-+\Gamma_H\right)\frac{H}{k_H}\nonumber\\
&&\qquad\,\,\,\,\,\,\, -\Gamma_Y\left[\frac{Q}{k_Q}-\frac{T}{k_T}+\left(\frac{H}{k_H}+\frac{h}{k_h}\right)\right]+S_{H}^{\CPV}\,,\nonumber\\
&&\partial^{\mu}h_{\mu}=-\Gamma_Y\left[\left(\frac{H}{k_H}+\frac{h}{k_h}\right)+\frac{Q}{k_Q}-\frac{T}{k_T}\right]\,.\label{eq:Boltzmannfullset}
\end{eqnarray}
\end{small}
The finite-temperature expressions for the coefficients $k_i$ can be found in Eq.  (72) of Ref.~\cite{Lee:2004we}\,. For reference, we note that in the massless limit $k_T=k_Q/2=k_B=3$ and $k_H=k_h=4$\,. Finite-temperature contributions to the thermal masses lead to modifications of these relationships.

\subsection{Approximate solution}
For purposes of deriving intuition about the transport dynamics, it is helpful to proceed toward a solution to Eqs.~(\ref{eq:Boltzmannfullset}) as far as possible analytically. To that end, it is useful to consider the limit  $\Gamma_Y, \Gamma_{ss}\rightarrow~\infty$, (see for example \cite{Lee:2004we} and the references therein). In this limit the linear combination of densities that multiply $\Gamma_Y$ and $\Gamma_{ss}$ relax to zero very close to the bubble wall, yielding two conditions that allow us to eliminate two of the four densities. We note that one difference with \cite{Lee:2004we} is that in our application we keep the $h$ density in our coupled equations, thus when we use the approximation of large $\Gamma_{Y}, \Gamma_{ss}$ we find 
\begin{eqnarray}
 &&\frac{T}{k_T}-\frac{Q}{k_Q}-\frac{H+h}{k_H}\approx 0,\label{eq:approx1}\\
 &&\frac{2Q}{k_Q}-\frac{T}{k_T}+\frac{9(Q+T)}{k_B}\approx 0\,.\label{eq:approx2}
\end{eqnarray}
Our four equations reduce to a set of two coupled Boltzmann equations which we solve numerically. For that reason we call this {\it{approximate solution}} as opposed to {\it{analytical solution}} which we would be able to achieve in the absence of the $h$ density. Using the approximate formulas \eq{eq:approx1} and \eq{eq:approx2} we solve for $T, Q$. The solution reads
\begin{eqnarray}
T=c_T\,(H+h),\qquad Q=c_Q\,(H+h)\,, \label{eq:cTcQequations}
\end{eqnarray}
where
\begin{eqnarray}
&&c_T=\frac{k_T(2k_B+9k_Q)}{k_H[k_B+9(k_Q+k_T)]},\nonumber\\
&&c_Q=\frac{k_Q(k_B-9k_T)}{k_H[k_B+9(k_Q+k_T)]}\,.  \label{eq:cTcQequations2}
\end{eqnarray}
The remaining two densities $H, h$ satisfy a set of two coupled Boltzmann equations. In order to find these equations we need to find linear combinations of equations in \eq{eq:Boltzmannfullset} that are free from either $\Gamma_{Y}, \Gamma_{ss}$\,. One possibility is to choose $2\partial_{\mu} T^{\mu}+\partial_{\mu} Q^{\mu}+\partial_{\mu}h^{\mu}$ and $\partial_{\mu}H^{\mu}-\partial_{\mu}h^{\mu}$. The resulting equations should be solved with respect to $H'', h''$ and as a result we obtain \footnote{We also neglect $\Gamma_{M}^+$\,. See the Table~\ref{tab:benchmarksABdata} for the validity of this approximation for our benchmark points $A, B$} 
\begin{eqnarray}
&&D_H\,H''-\left(a_{11}H'+a_{12}h'\right)-\left(\bar\Gamma_{11}H+\bar\Gamma_{12}h\right)+S_{1}=0,\nonumber\\
&&D_H\,h''-\left(a_{21}H'+a_{22}h'\right)-\left(\bar\Gamma_{21}H+\bar\Gamma_{22}h\right)+S_{2}=0,\nonumber\\ \label{eq:approxcoupledset}\end{eqnarray}
where we made the following definitions
\begin{eqnarray}
&&a_{11}=a_{22}=v_w\frac{D_H+(D_H+D_Q)(c_Q+2c_T)}{\bar{D}},\nonumber\\
&&a_{12}=a_{21}=v_w\frac{(c_Q+2c_T)(D_H-D_Q)}{\bar{D}},\nonumber\\
&&\bar\Gamma_{11}=\frac{[D_H+(c_Q+2c_T)D_Q](\Gamma_{M^-}+\Gamma_H)}{k_H\bar{D}},\nonumber\\
&&\bar\Gamma_{21}=-\frac{(c_Q+2c_T)D_Q(\Gamma_{M^-}+\Gamma_H)}{k_H\bar{D}},\nonumber\\
&&\bar\Gamma_{12}=\bar\Gamma_{22}=0\,,\nonumber\\
&&S_1=\frac{D_H+(c_Q+2c_T)D_Q}{\bar{D}}\,S_H^{\CPV},\nonumber\\
&& S_2=-\frac{(c_Q+2c_T)D_Q}{\bar{D}}\,S_H^{\CPV}\,,\nonumber\\
&& \bar{D}=D_H+2(c_Q+2c_T)D_Q\,.\label{eq:approximateformulaentries}
\end{eqnarray}
The left handed quark density is found via $n_L=4T+5Q$  \cite{Huet:1995sh}. The electroweak sphalerons transfer this density into the net baryon asymmetry according to \cite{Huet:1995sh}
\begin{eqnarray}
n_B=-3\frac{\Gamma_{\text{ws}}}{v_w}\int_{-\infty}^0 d z \,n_L(z)\exp\left({\frac{15}{4}\frac{\Gamma_{ws}}{v_w}}\right)\,. \label{eq:generatedbau}
\end{eqnarray}

Our expressions in \eq{eq:cTcQequations2} are equivalent to the $r_1$ term in an analogous approximate solution in the MSSM case as given in Eq. (84) in Ref.~\cite{Lee:2004we}. Note that if one uses the zero mass limit values of $k_{Q}=2k_T=2k_B$, from \eq{eq:generatedbau}, \eq{eq:cTcQequations}, \eq{eq:cTcQequations2} we see that the net baryon asymmetry equals to zero, as expected based on the vanishing of $r_1$ in this limit.  A non-vanishing $n_B$ then arises from retaining the subleading terms in $1/\Gamma_\mathrm{ss}$, as contained in the $r_2$-term in Eq. (84) of Ref.~\cite{Lee:2004we}. However, keeping the finite temperature contributions to the quark and scalar field masses yields a non-vanishing result at zeroth order in $1/\Gamma_\mathrm{ss}$. In Section \ref{sec:results} we compare the solution of \eq{eq:approxcoupledset} to the solution for the full set in \eq{eq:Boltzmannfullset}\,and we find for our benchmark scenarios that the approximate solution $Y_B$ is by $\sim10\%$ larger than the result of the full solution. We studied the agreement between the approximate and full method when varying the parameters of the theory in the wide range. In some points in the parameter space the approximate $Y_B$ becomes up to a factor of 2 larger than the full numerical solution, which cannot be explained by the subleading terms in $1/\Gamma_\mathrm{ss}$, but rather follows from the behavior of the densities in the vicinity of the bubble wall. Thus, we conclude that $Y_B$ is dominated by the leading order contribution in $1/\Gamma_\mathrm{ss}$ that results from retaining the finite-temperature masses in computing the statistical factors $k_j$.


\subsection{Profile functions}
The CPV source term in \eq{eq:sourcetermformula} is proportional to the combination of VEVs $v_{\Sigma}, v_S$ which is convenient to rewrite in the following way:
\begin{eqnarray}
&&v_S(x)v_{\Sigma}(x)\left[v_S(x)\dot{v}_{\Sigma}(x)-\dot{v}_S(x)v_{\Sigma}(x)\right]\nonumber\\
&&=\frac{1}{2}\,\left[v_{\Sigma S}(x)\right]^4\,\sin\,2\beta_{\Sigma S}(x)\,\dot{\beta}_{\Sigma S}(x)\,,\label{eq:sourcetermproportionaltothis}
\end{eqnarray}
where we have defined
\begin{eqnarray}
v_{\Sigma S}=\sqrt{v_{\Sigma}^2(x)+v_{S}^2(x)},\quad \beta_{\Sigma S}(x)=\arctan\frac{v_{\Sigma}(x)}{v_S(x)}\,.
\end{eqnarray}
The detailed shapes of the profiles $v_{\Sigma}, v_{S}$, or equivalently $v_{\Sigma S}, \beta_{\Sigma S}$ across the bubble are unknown and their detailed calculation is beyond the scope of this paper. Analogous calculation of the profiles in the MSSM have been performed in Ref.~\cite{Moreno:1998bq} and for the complex singlet extension in Ref.~\cite{Jiang:2015cwa}. In the MSSM case the role of two VEVs are playing $v_u(x), v_{d}(x)$. For simplicity we adopt the shape of the profiles from MSSM calculations, but stress that further study is required for a more complete treatment. Thus, we use the following analytical form \cite{Carena:1997gx,Carena:2000id,Lee:2004we,Cirigliano:2006wh} of the profile functions 
\begin{eqnarray}
&&v_{\Sigma S}(x)=\frac{1}{2}v^{(0)}_{\Sigma S}(T)\left[1+\tanh\left(\frac{2\alpha z}{L_w}\right)\right],\nonumber\\
&&\beta_{\Sigma S}(x)=\beta_{0}(T)-\frac{1}{2}\Delta \beta\left[1-\tanh\left(\frac{2\alpha z}{L_w}\right)\right]\,,\label{eq:VEVprofilefunctions}
\end{eqnarray}
with $\alpha=3/2$. Note that we have assumed the same wall thickness for the triplet and singlet vevs, an assumption that parallels the treatment of the doublet and singlet vevs in Ref.~\cite{Jiang:2015cwa}.

Similarly to the MSSM we will assume additional suppression of the BAU due to small numerical value of $\Delta \beta\sim 0.015$ \cite{Moreno:1998bq} . Unlike the MSSM case our source term is proportional not only to $\dot{\beta}(x)$, but also to $\sin 2\beta(x)$, which in principle could lead to additional suppression of the total generated BAU. For concreteness we choose the value for $\beta_0(T)$ in the early universe to be $\pi/4$, so that $\sin 2\beta_0(T)=1$\,. Our results can be appropriate re-scaled after a comprehensive study of the bubble profiles has been completed.

\section{Results}
\label{sec:results}
In this section we present the numerical results for bounds from existing and sensitivity regions from future generation EDM measurements, and also constraints from Higgs to diphoton decay and BAU. The question we are going to ask is whether it is possible to generate the observed BAU during the first step of the 2SPT described in this paper, in a way that is consistent with the experimental observations.

\subsection{EDMs}\label{subsec:ResultsEDMs}

\begin{figure*}[!t]
\includegraphics[scale=0.45]{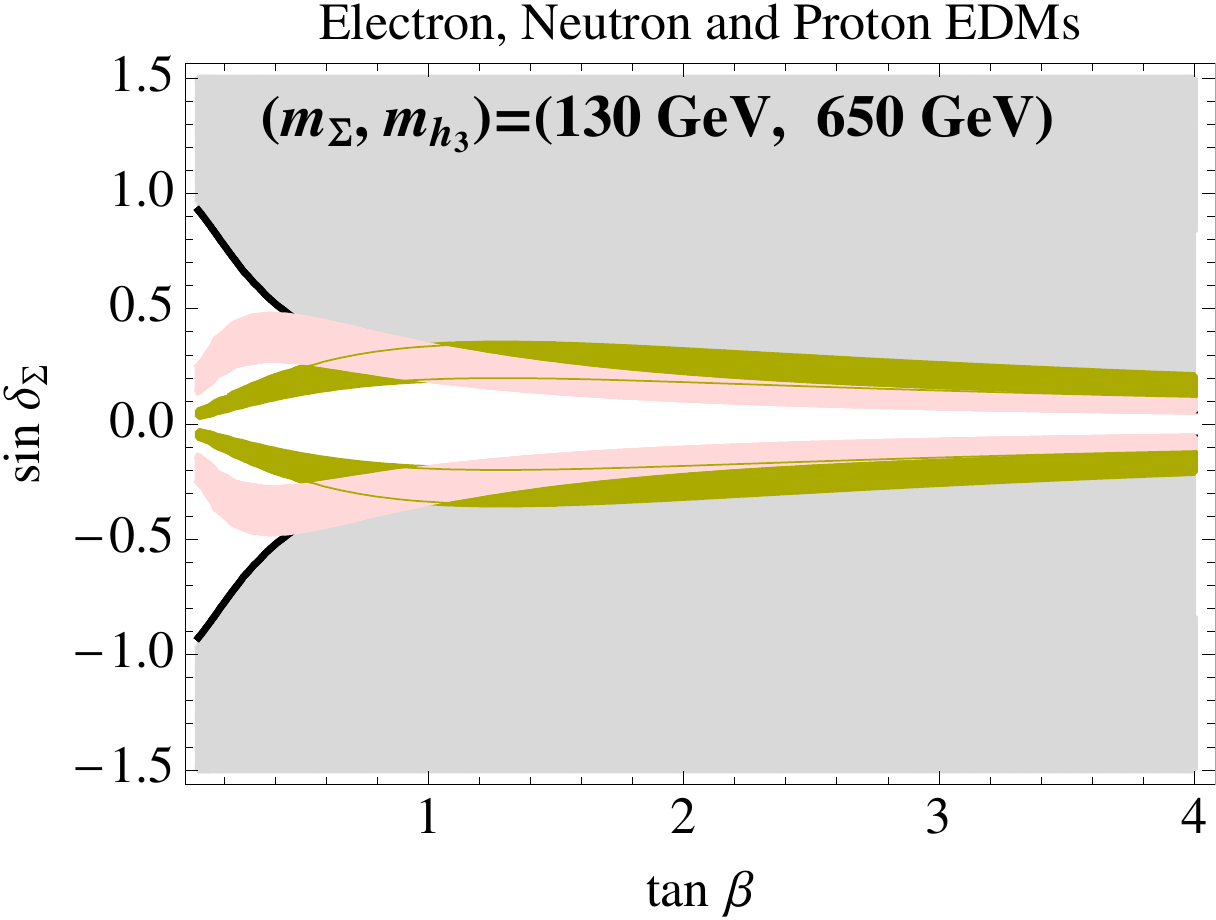}\quad \includegraphics[scale=0.45]{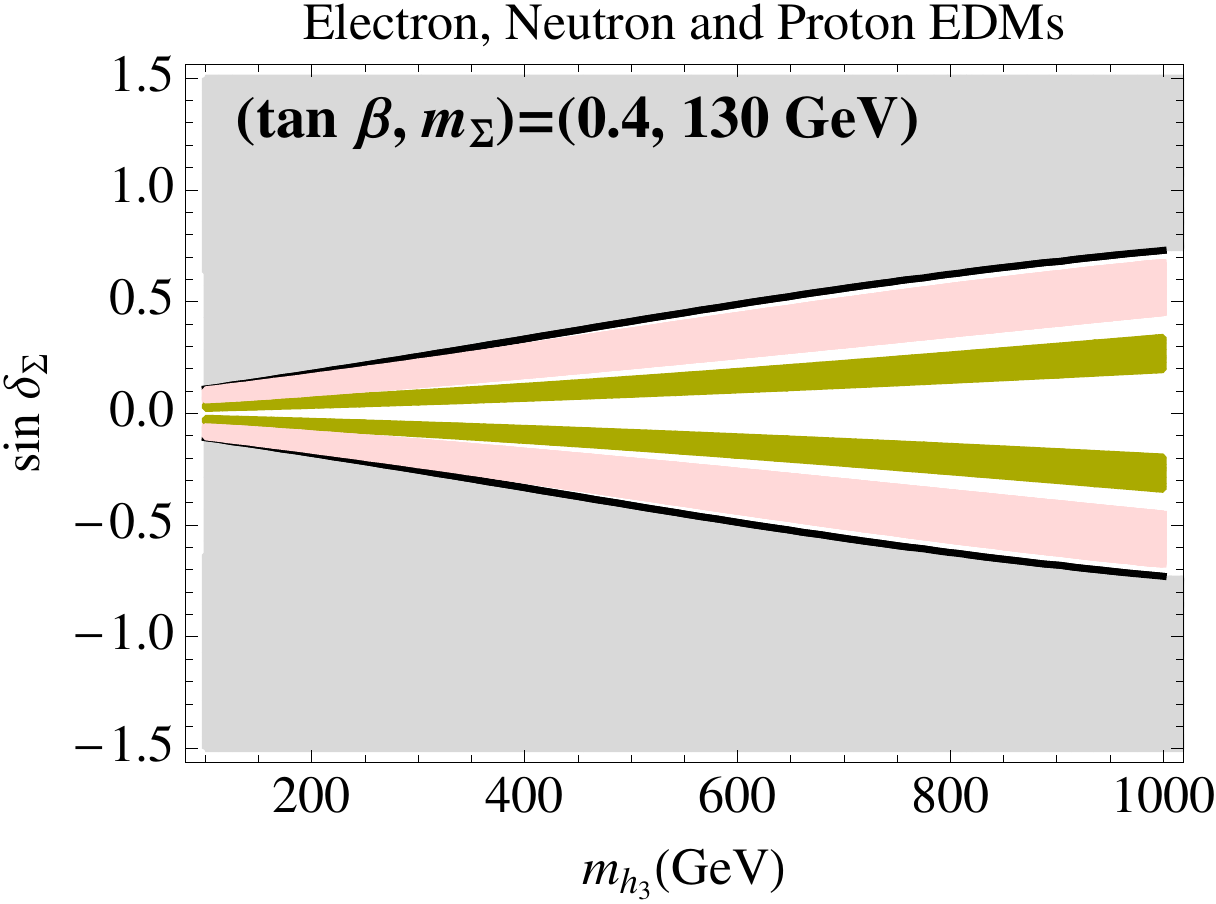}\quad\includegraphics[scale=0.45]{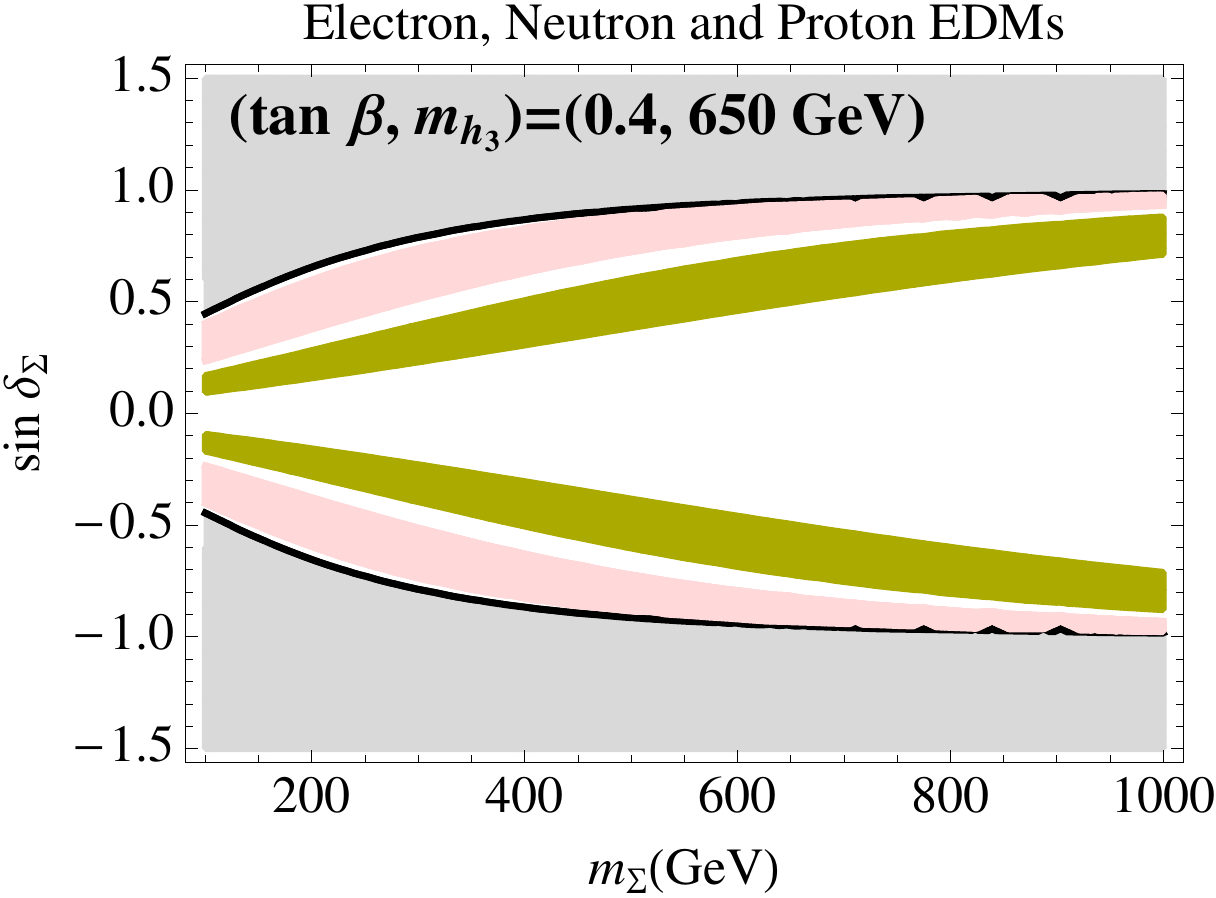}
\caption{Electron EDM bound $|d_e|<8.7\times10^{-29}e\text{\,cm}$ (gray band represents the excluded region), projected neutron sensitivity $|d_n|=2.9\times10^{-28}e\text{\,cm}$ (pink band) and a possible proton EDM with sensitivity  $|d_p|=2.0\times10^{-28}e\,\text{cm}$ (olive band). Widths of the pink and olive bands correspond to uncertainties in Eqs.~(\ref{eq:zetas}). See text for more details.}\label{fig:EDMbounds}
\end{figure*}

From our previous analysis the dependence of electron and neutron EDMs on the parameters of the theory has the following form
\begin{eqnarray}
&&d_e =\sin\delta_{\Sigma}\tan\beta\, F(m_{h_3},m_{\Sigma}),\label{eq:parametricdep}\\
&&d_n =\sin\delta_{\Sigma}\left(C^n_1\tan\beta+C^n_{2}\cot\beta\right)F(m_{h_3},m_{\Sigma})\,,\nonumber
\end{eqnarray}
where $C^n_1, C^n_2$ depend on the quark charges and nucleon matrix elements of up and down quark EDM operators. A similar expression to that for $d_n$ applies to the proton EDM with appropriate replacements $C^n_{1,2}\to C^p_{1,2}$. 

In Ref.~\cite{Bhattacharya:2015wna} the nucleon tensor charges have been evaluated on the lattice
\begin{eqnarray}
\rho_p^u=\rho_n^d=0.774(66),\qquad \rho_p^d=\rho_n^u=-0.233(28).
\end{eqnarray}
Using the above results we obtain from \eq{eq:edmmatrixelements} the following numerical values for the nucleon matrix elements of the quark EDM operators
\begin{eqnarray}
&&\zeta_{n}^{u}=(3.5\pm 1.0)\times 10^{-22}\text{\,cm},\nonumber\\
 &&\zeta_{n}^{d}=(-24.2\pm 2.9)\times 10^{-22}\text{\,cm},\nonumber\\
 &&\zeta_{p}^{u}=(-11.6\pm 3.2)\times 10^{-22}\text{\,cm},\nonumber\\  
 &&\zeta_{p}^{d}=(7.3\pm 2.1)\times 10^{-22}\text{\,cm}\,.
 \label{eq:zetas}
\end{eqnarray}
 Results based on sum rules and quark model can be found in the review \cite{Engel:2013lsa} and are approximately factor of 2 smaller with larger error bars. We have also used the PDG \cite{Agashe:2014kda} quark mass values $m_u=2.3\pm 0.6\,\text{MeV}$  and $m_d=~4.8\pm~0.4\,\text{MeV}$ \footnote{Note, that PDG  \cite{Agashe:2014kda} lists the light quark mass values $m_u=~2.3^{+0.7}_{-0.5}\,\text{MeV}$ and $m_d = 4.8^{+0.5}_{-0.3}\,\text{MeV}$, while we for simplicity have symmetrized the uncertainty on the quark masses.} and combined the relative uncertainties from quark masses and tensor charges in quadrature, assuming they are uncorrelated. The resulting uncertainties in  \eq{eq:zetas} range from 12\% to 29\%. Note, however that we have neglected possible contributions from the strange quarks, whose contribution to $d_n$ may be as large as 35\% with an uncertainty of similar magnitude\cite{Bhattacharya:2015wna}. Consequently, to be conservative, we will take the uncertainty on $d_{n/p}$ to be 29\%.


The constraints on the CPV phase $\delta_{\Sigma}$ are shown in Figure~\ref{fig:EDMbounds} as a function of $\tan\beta, m_{h_3}, m_{\Sigma}$\,. The gray bands correspond to the current ACME electron EDM bound $|d_e|<8.7\times 10^{-29}e\text{\,cm}$ at  the 90\% confidence level. The light-red bands correspond to the projected neutron EDM sensitivity $|d_{n}|=2.9\times 10^{-28}e\text{\,cm}$, which is a factor of 100 times more sensitive than the current neutron EDM bound. The dark yellow bands correspond to a possible proton EDM search with sensitivity of $|d_p|=2.0 \times 10^{-28}e\text{\,cm}$\,.  

As expected for the first panel in the Figure~\ref{fig:EDMbounds}, the electron EDM bound on $\sin \delta_{\Sigma}$ is a hyperbola as a function of $\tan\beta$, the neutron EDM bound becomes stronger at both large and small values of $\tan\beta$, consistent with parametric dependence in \eq{eq:parametricdep}\,. The second and third panels reveal the dependence of function $F(m_{h_3}, m_{\Sigma})$ in \eq{eq:parametricdep} on its arguments. Also as expected, the bounds become weaker as one increases the mass of either scalar that enters the two-loop contributions. In all three plots we indicate in the top left the numerical values used for the remaining two parameters, besides those that are varied in the plot.

From the left panel, we see that  the existing electron EDM bound yields tight constraints on the angle $\delta_{\Sigma}$ except at small values of $\tan\beta$\,. A future EDM experiment with $\sim 10^{-28}e$ cm sensitivity would cover this low-$\tan\beta$ region.

In addition we stress that the EDM bounds do not constrain the second CPV phase $\delta_{S}$, see \eq{eq:deltaiphases}\,. We will exploit this feature below when we discuss the numerical results for constraints coming from the BAU.

\subsection{Higgs to diphoton}\label{sec:Higgsdiphoton}
Using the current LHC data from run I, the Higgs to diphoton signal is consistent with SM model and equals \cite{Khachatryan:2014ira,Aad:2014eha}
\begin{eqnarray}
&&\mu_{\text{CMS}}=1.14^{+0.26}_{-0.23},\\
&&\mu_{\text{ATLAS}}=1.17^{+0.27}_{-0.27}\,.
\end{eqnarray}
These agree quite well and in our numerical analysis we use the combined result \cite{Aad:2015zhl}
\begin{eqnarray}
\mu_{\text{LHC}}=1.15^{+0.28}_{-0.25}\,.
\end{eqnarray}
Note that the diphoton signal strength above corresponds to the production mechanism via gluon gluon fusion. Adding the corresponding vector boson fusion measurement $1.17^{+0.58}_{-0.53}$  \cite{Aad:2015zhl} will not change this result significantly because the error bar for it is a factor of 2 bigger. 

\begin{figure}[!h]
\includegraphics[scale=0.55]{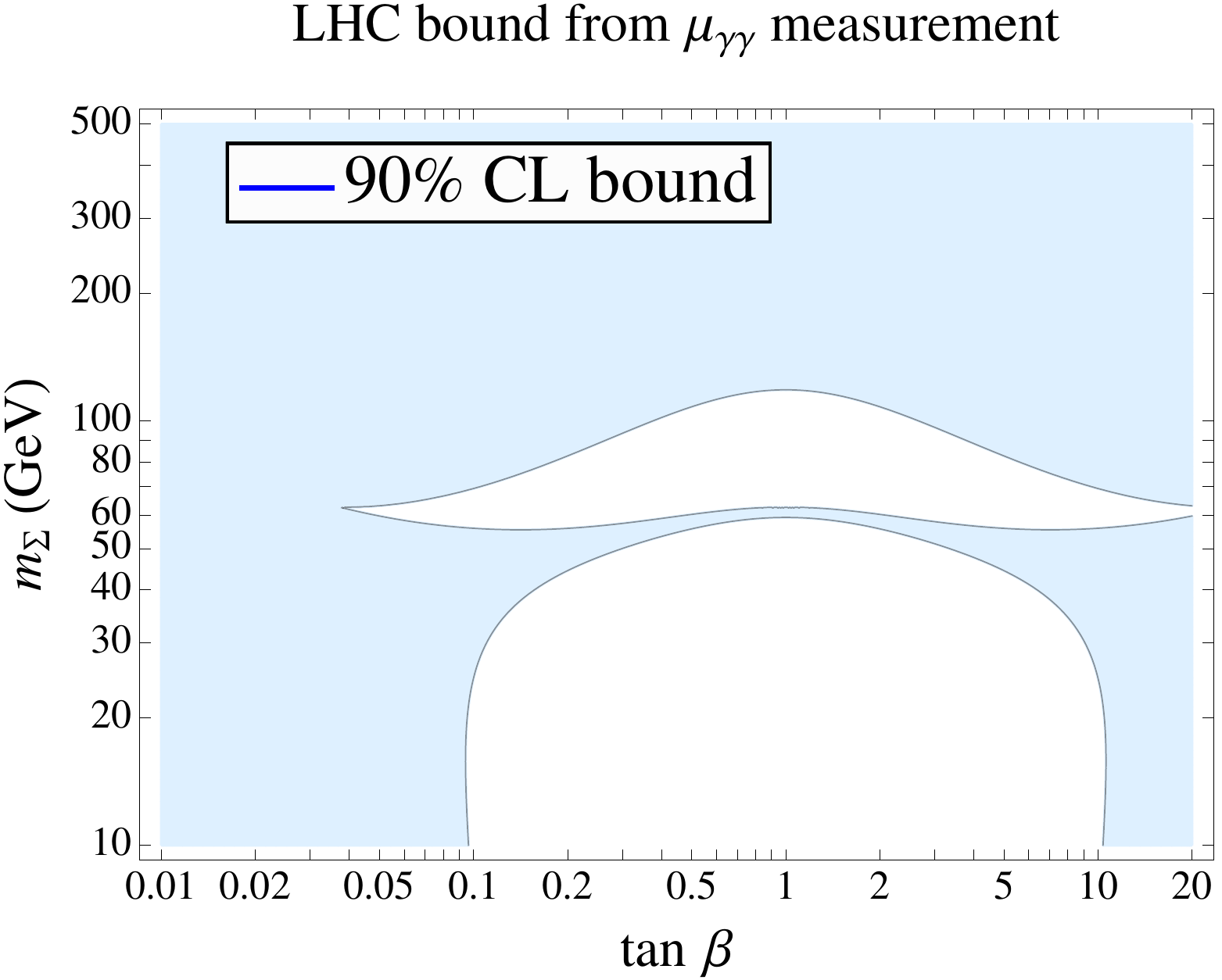}
\caption{\label{fig:Higgstodiphoton} Light-blue region represents the parameter space of the theory that at the 90\% confidence level is consistent with LHC diphoton data.}
\end{figure}

We see from equations \eq{eq:analysis1}-\eq{eq:analysis2} that all parameter-dependence of the Higss diphoton  rate can be absorbed into $m_{\Sigma}, g_{\Sigma}$, where the $g_{\Sigma}$ dependence on $\tan\beta, m_{\Sigma}, \text{Re}\,a_{2\Sigma}$ is given in \eq{eq:analysis2}\,.
The experimental constraints on the parameters $\tan\beta, m_{\Sigma}$ from the LHC diphoton data are shown in Figure~\ref{fig:Higgstodiphoton}. We have chosen $\text{Re} a_{2\Sigma}=1.07$ $-$ a value that is used in our BAU computation below. The light-blue shaded region corresponds to the parameter space of the theory consistent with the LHC diphoton data at 90$\%$ CL. 

From Figure~\ref{fig:Higgstodiphoton} we find that for $m_{\Sigma}>119\,\text{GeV}$ any value of $\tan\beta$ would in no tension with current LHC diphoton data. Thus we conclude that the run I LHC data on the Higgs  diphoton decay is only weakly constraining the parameters of our model; however we should keep in mind that run II data will place more stringent bounds.

\subsection{Generation of the baryon asymmetry}\label{eq:resultsbaryo}
In this subsection we will show that for illustrative vev profiles $v_{S}(x), v_{\Sigma}(x)$ adopted from MSSM studies and presented in \eq{eq:VEVprofilefunctions}, and for typical parameters for the bubble wall width ($L_w=0.25/T$) and wall velocity  ($v_w=0.05$), it is possible to generate the observed BAU during the first step of the 2SPT. For the critical temperature and the values of the vevs of the triplet and the singlet in the early universe we use  $T=123\text{\,GeV}, v_{\Sigma}(T)=v_S(T)=\text{76.3}\,\text{\,GeV}$, motivated by benchmark studies of \cite{Patel:2012pi}. For diffusion constants we use  $D_Q=6/T, D_H=110/T$ \cite{Chung:2009qs}. Again, a more comprehensive study would require an explicit analysis of a more appropriate choice of profiles and bubble wall parameters needed for the scenario at hand. 

\begin{figure}
\includegraphics[scale=0.345]{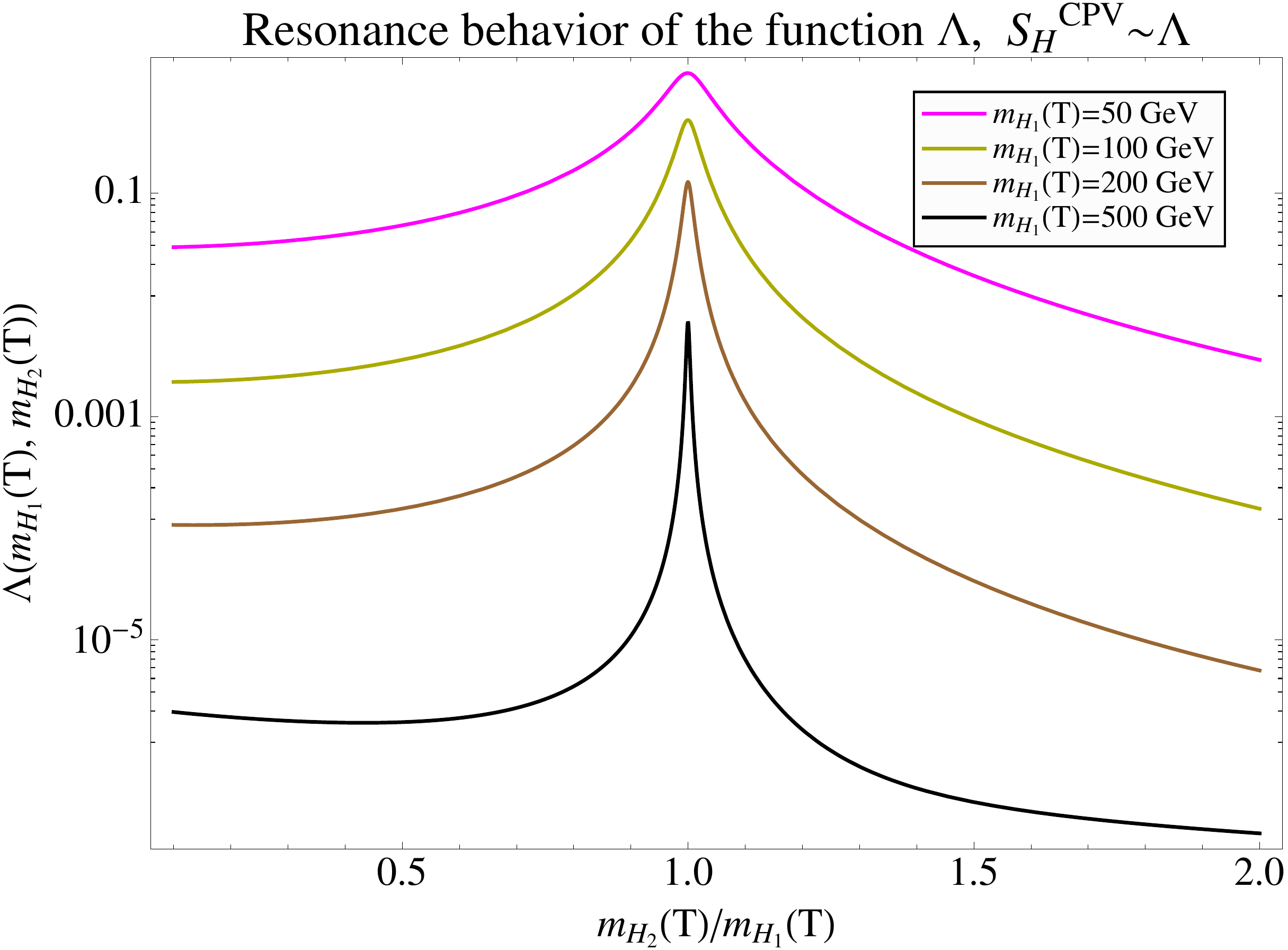}
\caption{Resonant behavior of BAU dependence on the parameters of the theory. See text for more details.}\label{fig:BaryoResonancePlot}
\end{figure}

The BAU has a known resonant behavior \cite{Lee:2004we}, reflected by a peak around some point in the parameter space. To see why this is the case, consider the CPV source term $S^{\CPV}_H$, which for our problem is shown in \eq{eq:sourcetermformula}\,. The function $\Lambda$ to which the CPV source is proportional to is given in \eq{eq:Wlambdaformulas}, and plotted in Figure~\ref{fig:BaryoResonancePlot}. The resonance leading to the maximum magnitude for the CPV source corresponds degeneracy of the thermal masses of the two Higgs bosons  $m_{H_1}(T)=m_{H_2}(T)$\,. It should be noted that similar resonant behavior is valid for the relaxation rates $\Gamma_{M}$ in \eq{eq:GammaM}, which to a certain degree mitigates resonance in the source\cite{Lee:2004we}.  The difference of mass squares of two Higgs bosons at finite temperature equals
\begin{eqnarray}
&&m_{H_1}^2(T)-m_{H_2}^2(T)\nonumber\\
&&=\left[y_b^2-y_t^2+\frac{m_{h_2}^2}{v^2}\left(\tan^2\beta-\cot^2\beta\right)\right]\frac{T^2}{4}\,.\label{eq:thermalmassdifference}
\end{eqnarray}
The BAU is, thus, maximal when  $\tan\beta$ and $m_{h_2}$ are chosen such that \eq{eq:thermalmassdifference} equals zero. Note that the Yukawa parameters $y_t, y_b$ also depend on $\tan\beta$ as indicated in the text below equation \eq{eq:LYFermion}.

As noted above, even though there exists no tree-level mixing between the two doublets due to the $Z_2$ symmetry of $V(H_1,H_2)$ ($m_{12}^2=0$), at the finite temperature, an off-diagonal mass term $\delta m^2(T)H_1^{\dagger} H_2+\text{h.c.}$ is generated with $\delta m^2(T)=\frac{a_{2\Sigma} T^2}{8}+\frac{a_{2S} T^2}{24}$. If non-negligible, such term would induce complicated flavor oscillations  \cite{Cirigliano:2009yt},  requiring proper treatment beyond the VIA. For consistency we estimate the error made by neglecting these flavor oscillations by defining parameter $\theta\sim \delta m^2(T)/(m_{H_1}^2-m_{H_2}^2)$ that characterizes the magnitude of flavor oscillations $P_{\text{osc}}\sim \theta^2$ \cite{Cirigliano:2009yt}. Therefore, if we choose to maximize the BAU and tune the parameters to be at the resonant point, the approximation of neglecting flavor oscillations becomes arbitrarily bad, or $\theta=\infty$\,. For practical purposes of selecting benchmark scenarios, one should select parameters away from the resonance, keeping the magnitude of $\theta$ from $\infty$ at reasonably controlled value. For the benchmarks that we study below the value of $|\theta|^2$ is included into the Table~\ref{tab:benchmarksABdata}\,.

\begin{table}
\caption{\label{tab:benchmarksABdata} Table of parameter values, thermal masses, relaxation rates, values of EDMs and BAU for benchmarks $A, B$.}
\begin{ruledtabular}
\begin{tabular}{lcccr}
Parameter &$A$& $B$\\
\hline
$m_{h_2}$  [GeV]&180&180\\
$m_{h_3}$  [GeV] & 650 & 650\\
$m_{\Sigma}$  [GeV] & 130 & 130\\
$m_{H^+}$  [GeV] & 300&300\\
$\tan\beta$ & 0.4 &0.4\\
$\sin\delta_{\Sigma}$ & 0.38 & 0\\
$\sin\delta_{S}$ & 0 & -0.4\\
$\text{Re}\, a_{2\Sigma}$ & 1.07 &1.07\\
$|a_{2S}|$ & 2.0 &2.0\\
$b_4$ & 0.8 &0.8\\
\hline
$m_{q_L}(T)$ [GeV] & $106$ &$106$ \\
$m_{t_R}(T)$ [GeV]& $133$ &$133$ \\
$m_{b_R}(T)$ [GeV] & $63$ & $63$\\
$m_{H_1}(T)$ [GeV]&  $82$ & $82$\\
$m_{H_2}(T)$ [GeV] &$215$ & $215$ \\
$m_{\Sigma}(T)$ [GeV] & $91$& $91$\\
$m_{S}(T)$ [GeV] & $200$& $200$\\
$\Gamma_{H_1}$ [GeV] & $5$ & $5$\\
$\Gamma_{H_2}$  [GeV]& $5$ & $5$\\
$\Gamma_Y$ [GeV] & $1.9$ &$1.9$ \\
$\Gamma_H$ [GeV] & $0.31$ & $0.26$\\
$\Gamma_{ss}$ [GeV] & $0.41$ & $0.41$\\
$\Gamma_{M^+}$ [GeV] & $-0.084$ & $-0.080$ \\
$\Gamma_{M^-}$ [GeV] &$0.48$ & $0.46$ \\
\hline
$d_e/\left(10^{-29}e\,\text{cm}\right)$ & 5.8& $0$  \\
$d_n/\left(10^{-28}e\,\text{cm}\right)$ &-3.3 & $0$ \\
$d_p/\left(10^{-28}e\,\text{cm}\right)$ &6.0 & $0$ \\
$Y^{\text{approx}}_B/\left(10^{-11}\right)$ & 9.3 & 9.1\\
$Y^{\text{full num}}_B/\left(10^{-11}\right)$ &8.6 & 8.4 \\
$|\theta|^2$ &$0.0073$ & $0.0066$\\
$\mu_{\gamma\gamma}$ & 0.86& 0.86\\
\end{tabular}
\end{ruledtabular}
\end{table}
\begin{figure*}[!t]
\includegraphics[scale=0.35]{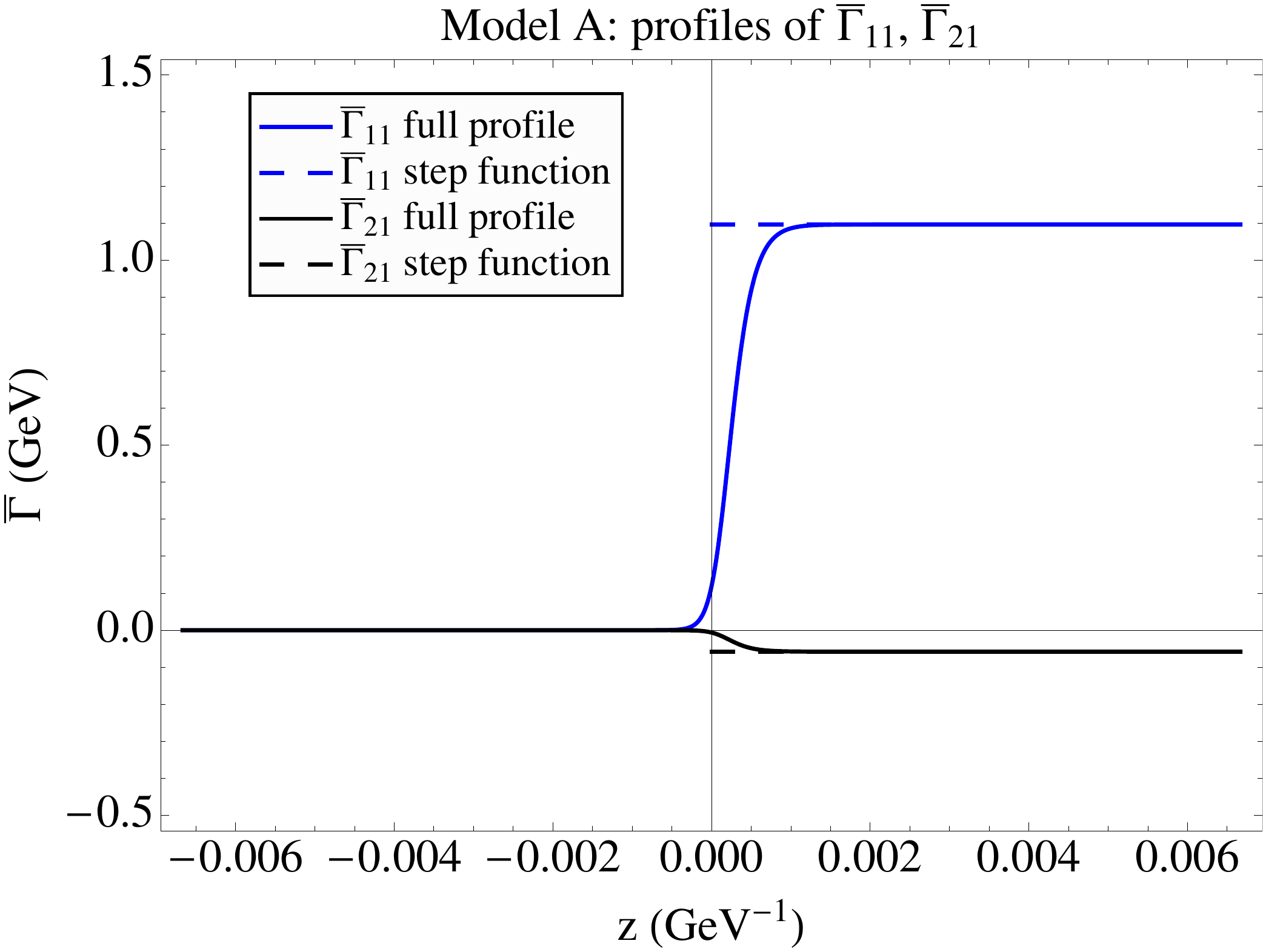}\qquad\includegraphics[scale=0.35]{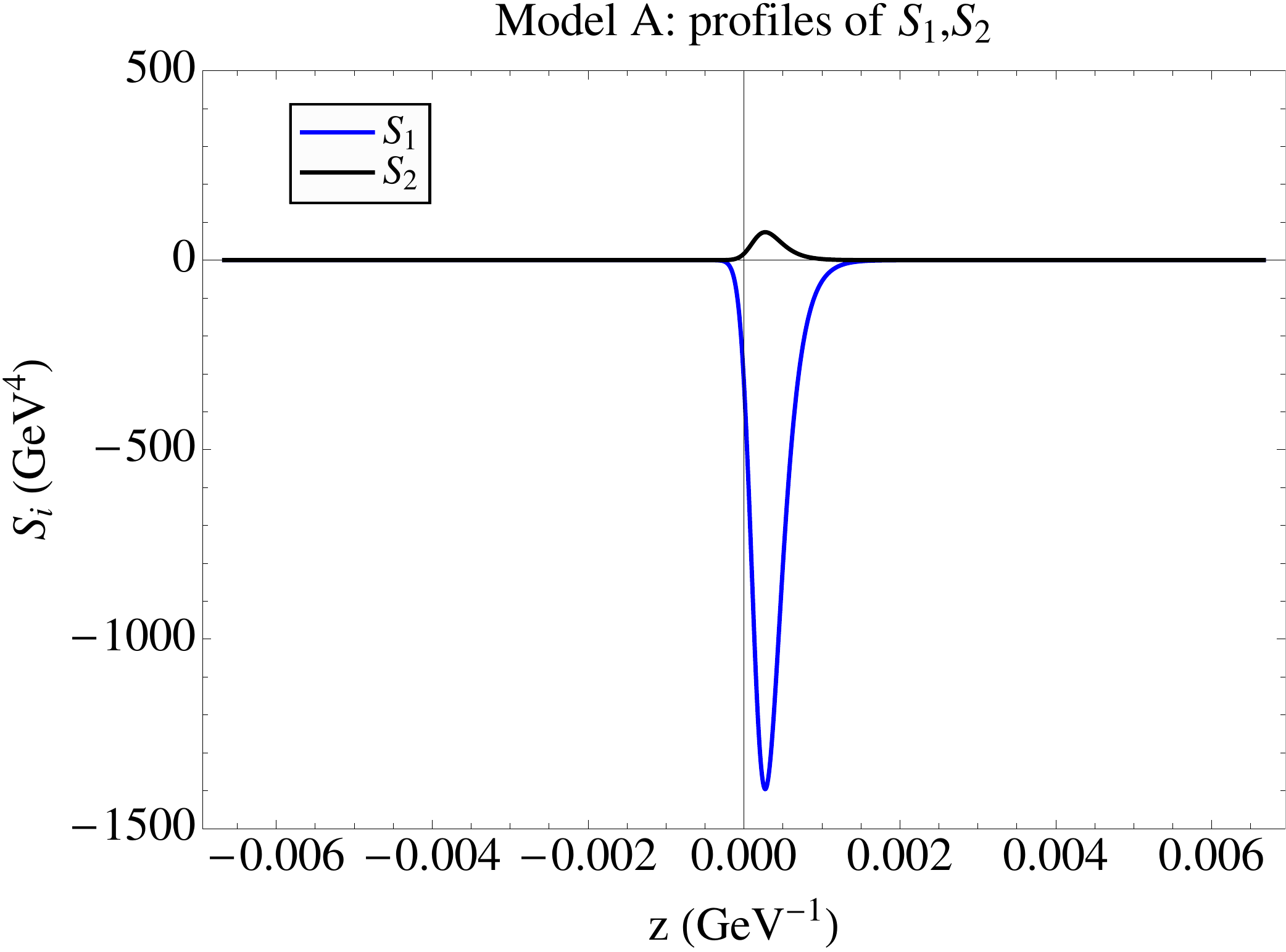}\\
\includegraphics[scale=0.35]{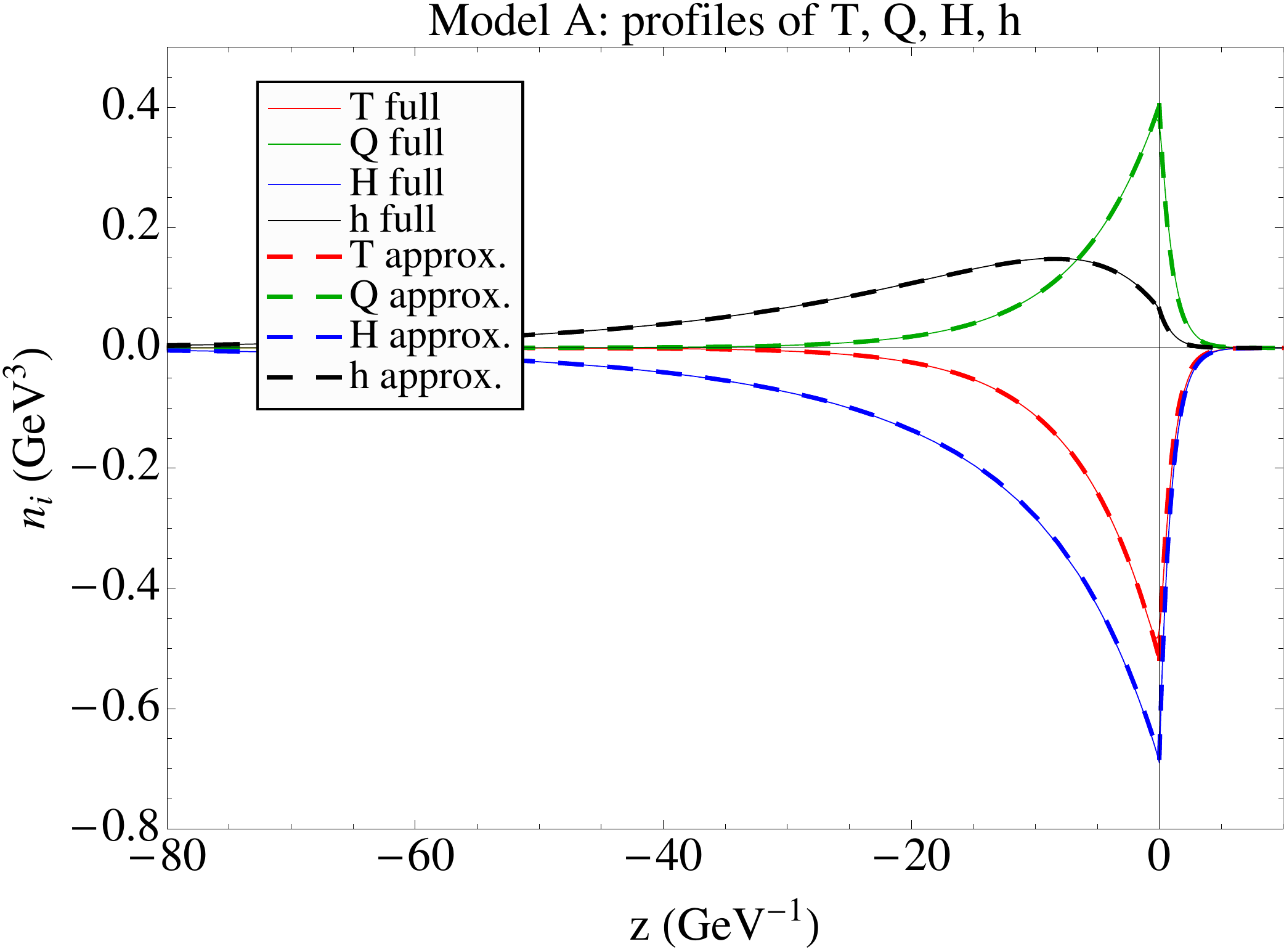}\qquad\includegraphics[scale=0.35]{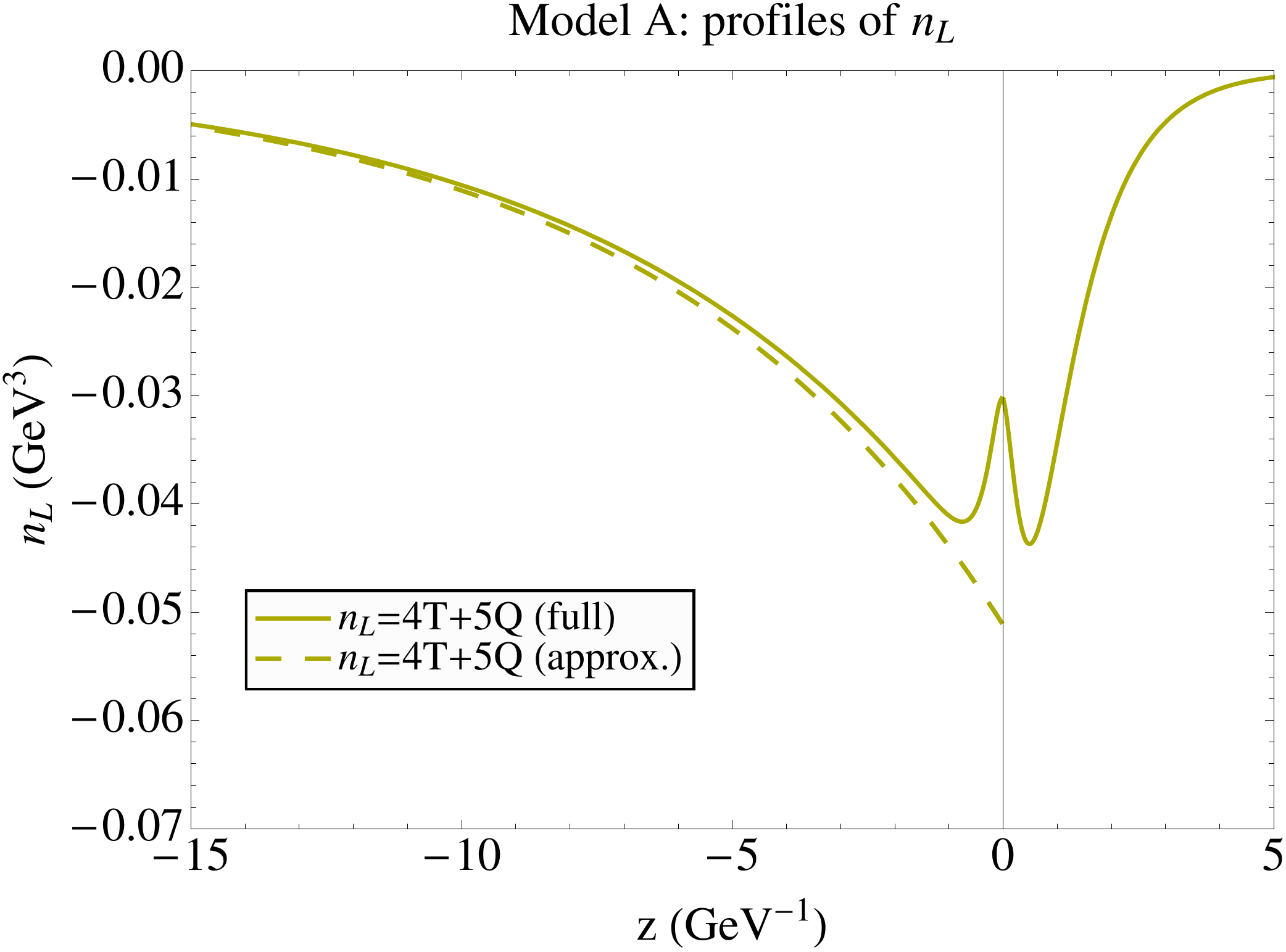}
\caption{Profile functions and the solution to the boundary problem for benchmark $A$. See text for more details.}\label{fig:BaryoModelABprofiles}
\end{figure*}

With these considerations in mind, we select two benchmark scenarios $A$ and $B$ that yield the observed BAU while respecting the present EDM bounds and avoiding significant flavor oscillations. We set $\delta_{S}=0$ for benchmark $A$ and $\delta_{\Sigma}=0$ for the benchmark $B$. The remaining parameters, along with the thermal masses, relaxation rates in the broken phase, and the values of the EDMs and BAU are summarized in the Table~\ref{tab:benchmarksABdata}\,\footnote{Note that our choice of parameters $\text{Re}\,a_{2\Sigma}, b_{4\Sigma}, T, v_{\Sigma}(T)$ is motivated by 2SPT benchmark studies of Ref.\cite{Patel:2012pi}.}. As one can see from that table, the neutron EDM is currently consistent with benchmark $A$, however with a factor of 100 increase in the sensitivity this benchmark will be probed and possibly ruled out. On the other hand benchmark present and future EDM searches have no sensitivity to the benchmark $B$.

The profiles of the relaxation rates, sources and particle-antiparticle asymmetries for benchmark point $A$ are illustrated in the Figure~\ref{fig:BaryoModelABprofiles}. The first panel represents the dependence of two effective relaxation rates $\bar{\Gamma}_{11}$ and $\bar{\Gamma}_{21}$ entering the approximate equations in \eq{eq:approxcoupledset}\, on the co-moving distance from the bubble wall, $z=x+v_w t$, with 
$z<0$ corresponding to the symmetric phase (outside of the bubble) and $z>0$  to the broken phase (inside the bubble). The second panel represents the two sources $S_1, S_2$ entering the approximate formula. Note the qualitative features that the relaxation rates $\bar\Gamma$ are zero in the symmetric phase and look like a step function across the bubble wall, while the CPV sources are zero everywhere except within the bubble wall. The smallness of the magnitude of the $S_2$ compared to $S_1$ can be understood from explicit formulae for them in \eq{eq:approximateformulaentries} and the fact that numerically $D_Q\ll D_H$.


The third panel represents all the four densities $T,Q,H,h$ in colors red, green, blue black respectively. The thin solid lines correspond to the full numerical solution to the transport equations while the thick dashed lines to the approximate method. Finally, the last panel represents the left handed quark density $n_L=4T+5Q$  \cite{Huet:1995sh}, which is converted to the net baryon asymmetry via electroweak sphalerons \cite{Huet:1995sh}, see \eq{eq:generatedbau}\,. Figure~\ref{fig:BaryoModelABprofiles} serves as an illustration how particle-antiparticle densities are distributed inside and outside of the bubble in the early Universe. The last panel also shows that the approximate method to solve the transport equations is in a reasonable agreement with the full numerical solution, except in the vicinity of the bubble wall. While the agreement between the full and approximate solutions in the third panel might lead one to expect similar agreement in the last panel, we note that the LH density $n_L=4T+5Q$ is subject to large cancellations. Consequently, any small differences between the full and approximate solutions in the third panel become magnified in $n_L$ (note the order of magnitude smaller vertical scale in the fourth panel as well).  Thus, it is not surprising that expectations based on the approximate solution are fairly well reflected in the final results for $Y_B$.
\begin{figure*}[!t]
\includegraphics[scale=0.51]{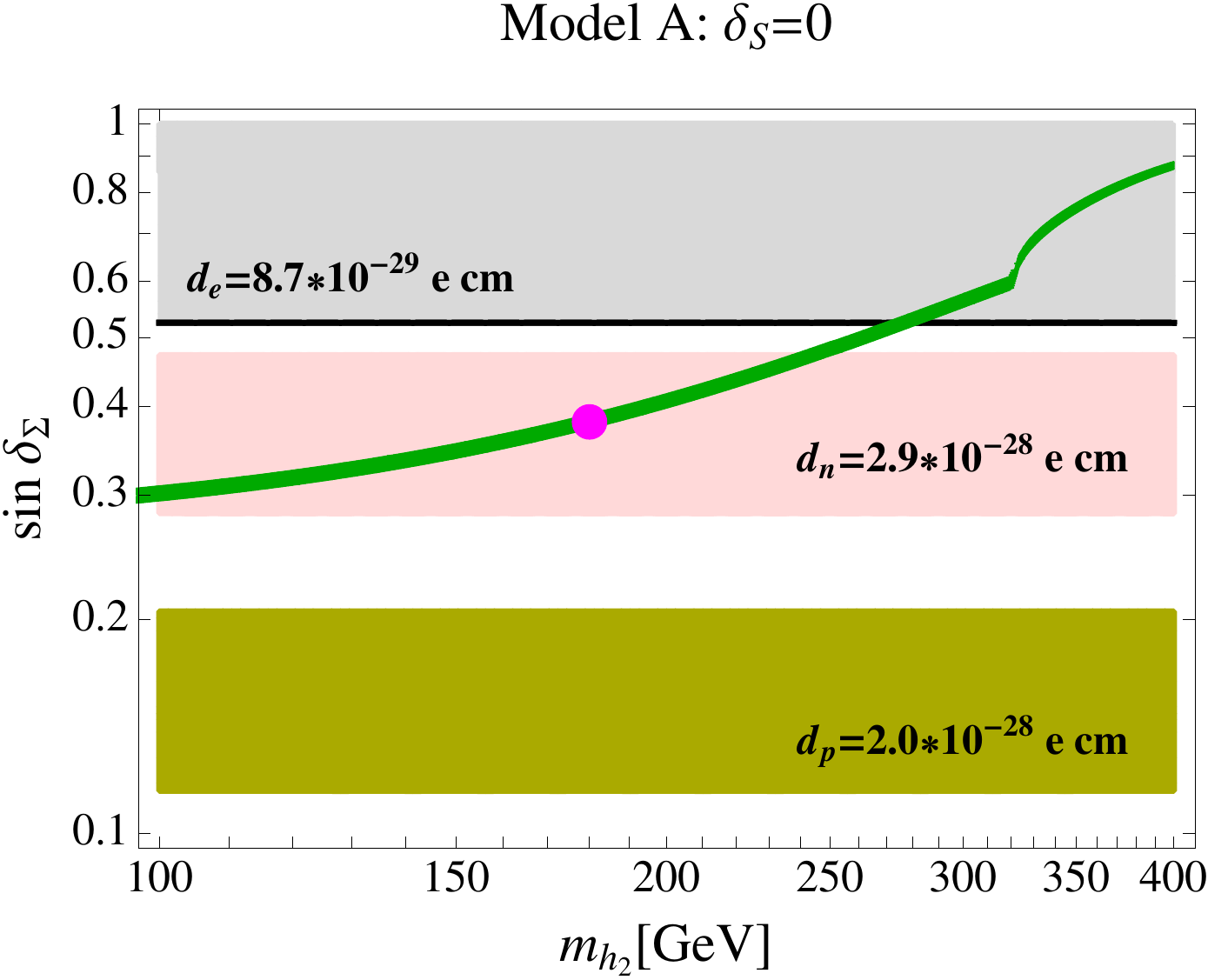}\qquad \includegraphics[scale=0.51]{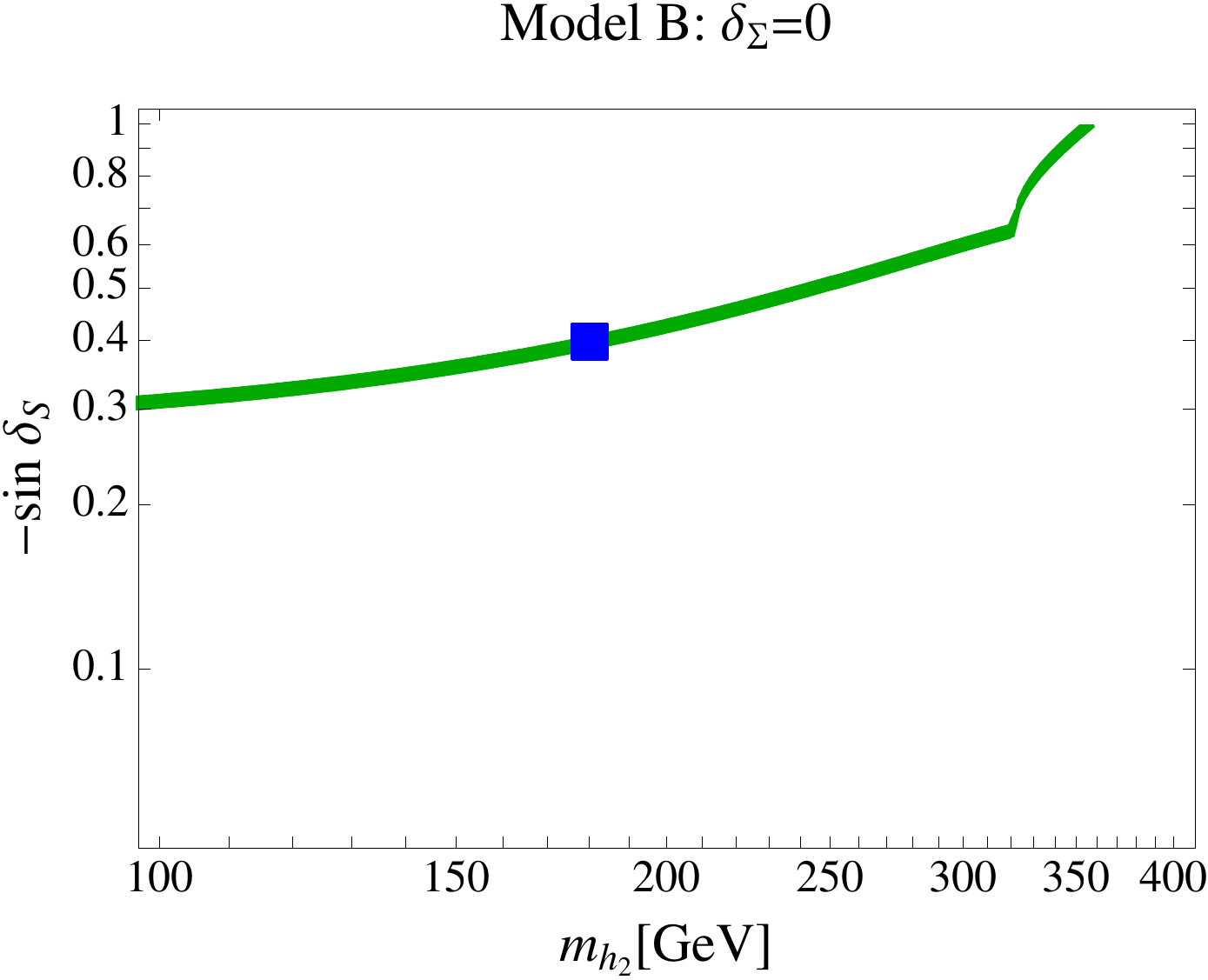}\\
\caption{ Constraints on the CPV phases $\delta_\Sigma$ (left panel) and $\delta_S$ as a function of the mass $m_{h_2}$. Solid green band is consistent with the observed BAU: $Y_{B}=(8.59\pm 0.22)\times 10^{-11}$.  For $\delta_S=0$ (left panel) the current electron EDM bound $|d_e|<8.7\times 10^{-29}e\,$~cm excludes the shaded region above the horizontal black line.  Sensitivities of a future neutron EDM $|d_n|<2.9\times 10^{-28}e\,$~cm\cite{Baron:2013eja} and possible proton EDM with $|d_p| < 2.9\times 10^{-28}e\,$~cm are indicated, respectively, by the pink and olive shaded horizontal bands. Widths of the bands correspond to uncertainties in Eqs.~(\ref{eq:zetas}). See text for more details. Present and future EDM searches have no sensitivity when $\delta_\Sigma=0$ (right panel). See Table~\ref{tab:benchmarksABdata}  parameters corresponding to benchmarks $A$ (magenta circle) and $B$ (blue square). For each panel, we fix these values and vary only the two parameters shown.}\label{fig:ModelsABmh2}
\end{figure*}


The resulting baryon asymmetry as a function of the CPV phases $\delta_\Sigma$ and $\delta_S$ and $m_{h_2}$ are shown in Fig.~\ref{fig:ModelsABmh2}. In the left (right) panel we set $\delta_S=0$ ($\delta_\Sigma=0$). We also indicate the present EDM constraints and prospective future sensitivities, which appear only in the left panel since for $\delta_\Sigma=0$ the interactions in $V(H_1,H_2, \Sigma, S)$ generate no elementary fermion EDMs through two-loop order. The present electron EDM bound $|d_e|<8.7\times 10^{-29}e\,$~cm obtained by the ACME collaboration\cite{Baron:2013eja} excludes the shaded region above the horizontal black line. The horizontal pink band indicates the reach of the future neutron EDM search underway at the Fundamental Neutron Physics Beamline at the Spallation Neutron Source that has a goal sensitivity of $|d_{n}|=2.9\times 10^{-28}e\text{\,cm}$. A possible future proton EDM search with a sensitivity of $|d_{p}|\sim 10^{-29}e\text{\,cm}$ would cover the entire BAU-viable region of the left panel. For illustrative purposes, we show the reach with a proton EDM experiment having $|d_{p}|= 2.0\times 10^{-28}e\text{\,cm}$ sensitivity with the olive band. The widths of the proton and neutron EDM bands correspond to 29\% error, as in Fig.~\ref{fig:EDMbounds}. The green bands correspond to the parameters for which the observed baryon asymmetry is generated. The discontinuity in the slope at $m_{h2}\sim 350$ GeV results from crossing a kinematic threshold in the three-body, particle number changing rates.

These results indicate that it is possible for the observed BAU to be generated during the first step of the two-step EWPT. The present electron EDM bound excludes a portion of the BAU-viable parameter space associated with the CPV phase $\delta_\Sigma$, while future nucleon EDM searches could probe most or even all of this sector of the model. On the other hand, the source of the BAU associated with the singlet-Higgs operator $a_{2S} H_1^\dag H_2 S^2+\mathrm{h.c.}$ is immune from these present and future EDM probes. 

\begin{figure*}[!t]
\includegraphics[scale=0.55]{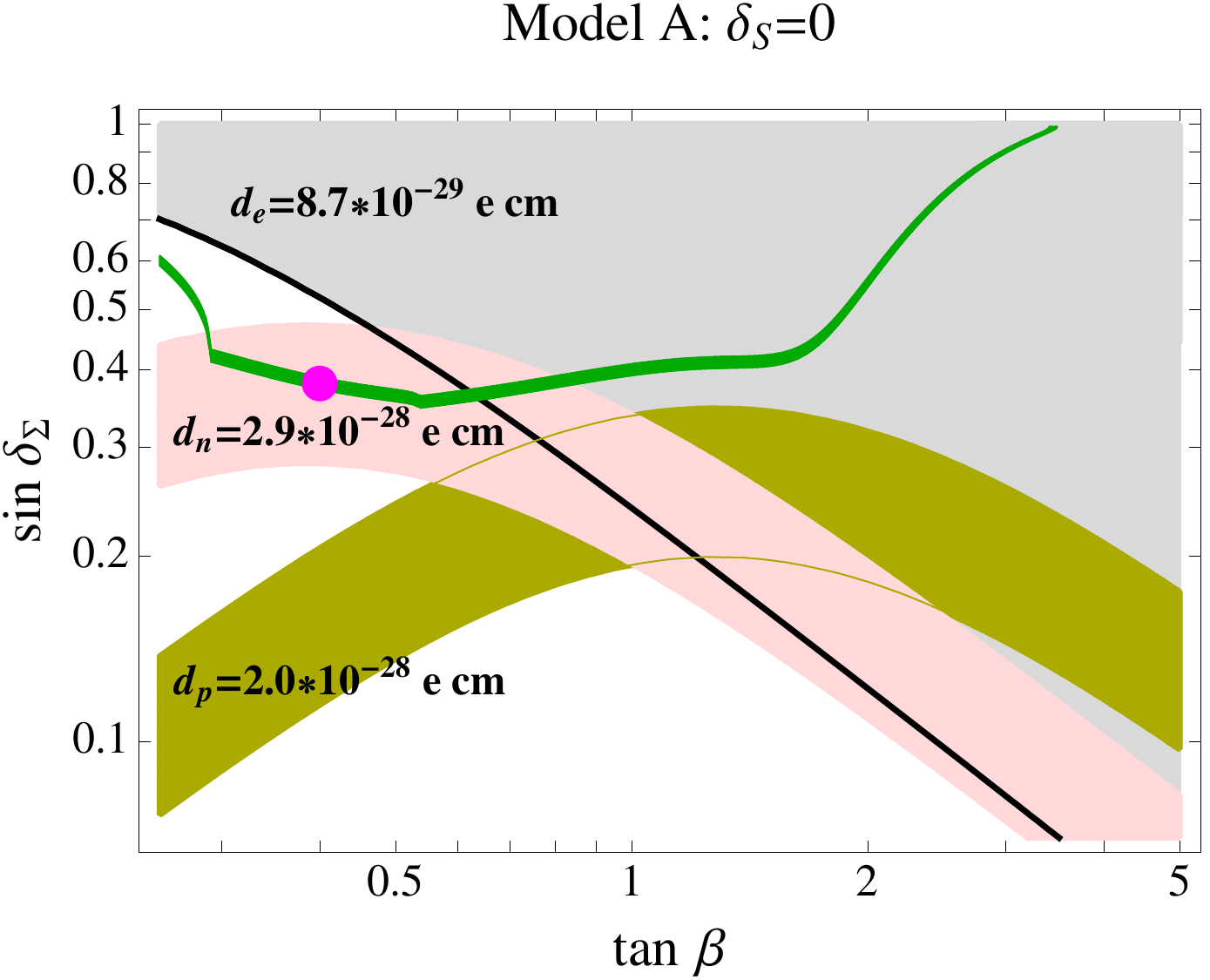}\qquad \includegraphics[scale=0.55]{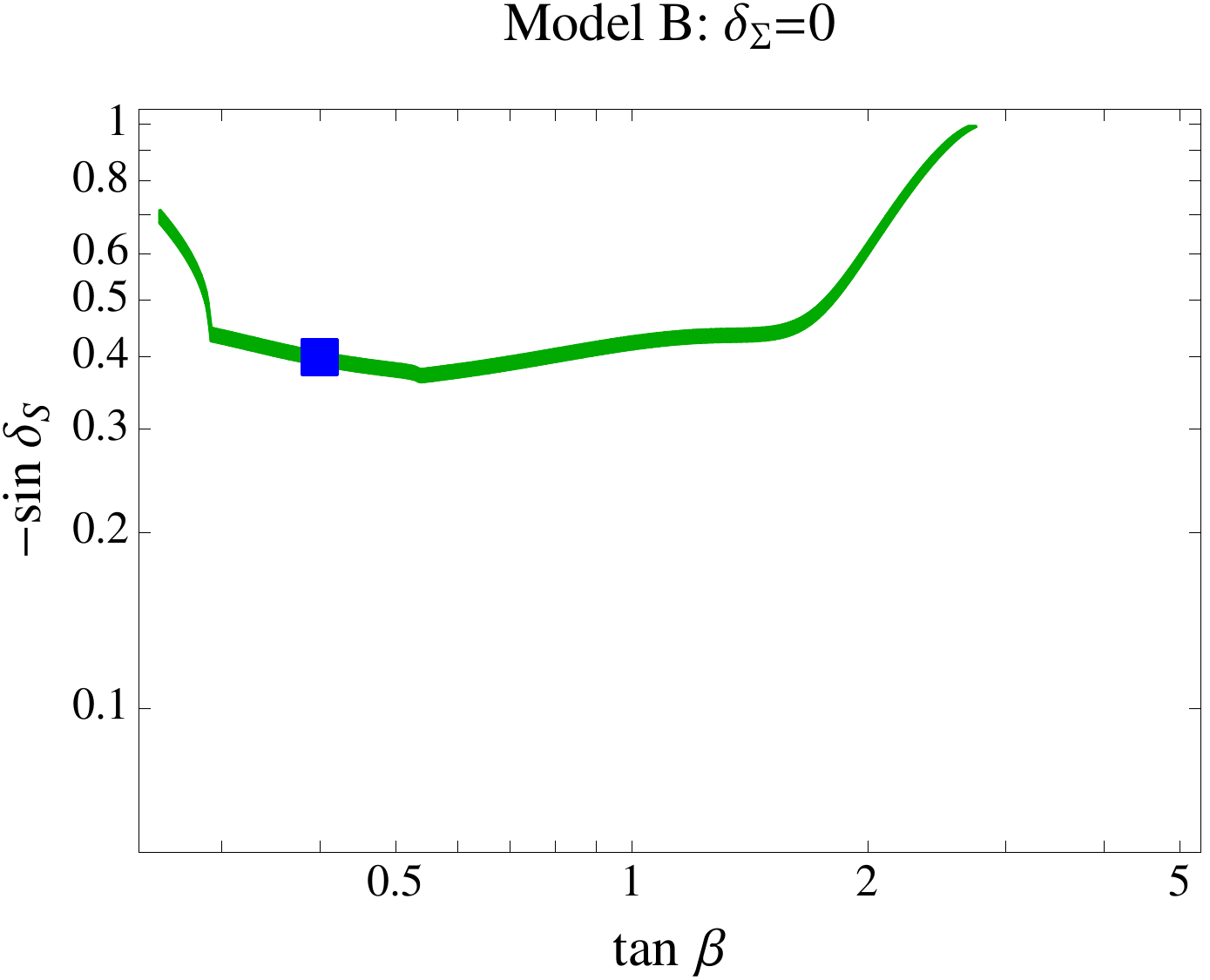}\\
\caption{ Constraints on the CPV phases $\delta_\Sigma$ (left panel) and $\delta_S$ as a function of the $\tan\beta$ for fixed $m_{h_2}=180$ GeV. Various regions and curves have the same meaning as in Fig.~\ref{fig:ModelsABmh2}.
}\label{fig:ModelsABtanbeta}
\end{figure*}

It is also interesting to ask how these statements vary with the other parameters in the theory, particularly $\tan\beta$. To that end, we show in Fig.~\ref{fig:ModelsABtanbeta} the $\tan\beta$-dependence of the CPV phases and EDM sensitivities. We restrict our consideration to $\tan\beta\lsim 5$, as for values above this region, one must include explicitly the effects of bottom quark Yukawa rates that are enhanced as $\tan^2\beta$\cite{Chung:2008aya} in the type II 2HDM. Note that for Model A, the present electron EDM bound restricts one to values of $\tan\beta$ below unity.  For Model B, where the EDM places no constraint on the CPV phase $\delta_S$, we observe that consistency with the observed BAU requires $\tan\beta\lsim 3$. Based on experience with the MSSM, it is likely that in the large $\tan\beta$ regime for both models there may be significant cancellations between effects associated with top and bottom Yukawa rates, leading to an even smaller $Y_B$ and the requirement of larger CPV phases. Consequently, we expect that this regime will not be viable, and we defer a detailed study of this regime, as well as an analysis of the type I 2HDM realization of our model, to future work.

In principle, it would also be interesting to explore the $m_\Sigma$-dependence. In the present set-up, larger values of $m_\Sigma$ would not be consistent with the two-step EWSB scenario according to the analysis of Ref.~\cite{Patel:2012pi}. For smaller values of $m_\Sigma$, the EDMs induced by $\delta_\Sigma$ become larger, leading to ever more severe constraints for Model A. 
We observe that allowing (a) the coefficient of the $\Sigma^2 S^2$ operator to be non-vanishing and (b) $\langle S\rangle \not=0$ at $T=0$ would open the possibility of larger values of the $T=0$ triplet mass, thereby in principle weakening the EDM constraints on Model A, while preserving the viability of the two-step EWSB transition at finite $T$. We again defer a detailed study of this possibility, including the impact on asymmetry generation, to future work.

\section{Conclusions}
\label{sec:conclude}
The possibility that EWSB occurred in multiple steps opens a new pathway for weak scale baryogenesis. In the two-step EWBG paradigm, the BAU is generated by CPV dynamics during a first order transition to an EWSB-vacuum that precedes the final transition to the Higgs phase of today's universe. As outlined in Section \ref{sec:general}, there exist multiple possibilities for the CPV interactions that could make this new pathway effective. In this study, following the previous work in Ref.~\cite{Patel:2012pi}, we have illustrated one subset of these possibilities, focusing on renormalizable interactions in the scalar sector that directly generate CPV-asymmetries in the SM Higgs sector via Higgs portal interactions. In this example, BAU generation requires the presence of two fields that obtain space-time varying vevs during the first EWSB step and two fields in the Higgs sector that mix due to CPV-interactions with one or both of these vevs. For the former, we choose a real triplet ${\vec\Sigma}$, whose vev breaks EW symmetry during the first step, and a real singlet. For the latter, we employ a type II 2HDM, wherein neither of the neutral doublet fields obtain vevs during the first step but do so in the second step (see Fig.~\ref{fig:TwoStepIllustration}). The real singlet provides an example of a \lq\lq partially excluded" sector that interacts with the SM solely via the new CPV interaction. 

Several generalizable features emerge from our illustrative model study. Most significantly, the two-step EWGB paradigm appears to be a viable mechanism for creating the BAU and does not appear to require fine tuning of parameters. Moreover, while a portion of the CPV dynamics is accessible to present and future EDM searches, those associated with the partially secluded sector are not. Direct searches for the new scalar states could either discover exclude the  ingredients necessary for this scenario, but direct tests of the CPV interactions are limited to those involving non-singlet fields, at least for the foreseeable future. On an experimentally more positive note, should the CPV responsible for the BAU involve the non-secluded sector, one could anticipate non-vanishing signals in future EDM searches.

Looking ahead, it would be interesting to explore both other specific realizations of two-step EWBG as well as to study the present example with greater comprehensiveness. The latter analysis would include consideration of the type I 2HDM; allowing for a non-vanishing $a_{2\Sigma S}$ coupling that could yield larger triplet masses consistent with the strong first order EWPT during the first step; carrying out a detailed study of the bubble profiles and wall velocities; and ultimately going beyond the VIA along the lines of Refs~\cite{Cirigliano:2009yt,Cirigliano:2011di}.

{\bf{Acknowledgements. }}It is pleasure to thank Wei Chao, Vincenzo Cirigliano, Christopher Lee, David Morrissey, Sean Tulin and Peter Winslow for many useful discussions. This work was supported in part by
U.S. Department of Energy contract DE-SC0011095.

\appendix
\section{EDM loop integrals}\label{sec:appendixEDM}
Here we summarize the loop functions needed for the EDM calculation.
\begin{eqnarray}
&&f(z)=\frac{z}{2}\int_0^1 d x\frac{\left[1-2x(1-x)\right]\ln\frac{x(1-x)}{z}}{x(1-x)-z},\nonumber\\
&&g(z)=\frac{z}{2}\int_0^1 d x\frac{\ln\frac{x(1-x)}{z}}{x(1-x)-z},\nonumber\\
&&\tilde{f}(x,y)=\frac{y f(x)}{y-x}+\frac{x f(y)}{x-y},\nonumber\\
&&\tilde{g}(x,y)=\frac{y g(x)}{y-x}+\frac{x g(y)}{x-y}.
\end{eqnarray}
\section{Thermal masses}\label{sec:appendixBaryogenesis}
Thermal masses for the fields $q_L,t_R,b_R, H_1, H_2,\Sigma,S$ are summarized in the Table~\ref{tab:thermal_masses_appendix}.
\begin{table}
\caption{\label{tab:thermal_masses_appendix} Thermal masses. }
\begin{ruledtabular}
\begin{tabular}{lcr}
Field&Thermal mass $\delta m_{\text{SM}}^2/T^2$\\
\hline
$q_L$ & $\frac{1}{6}g_3^2+\frac{3}{32}g_2^2+\frac{1}{288}g_1^2+\frac{1}{16}y_t^2+\frac{1}{16}y_b^2$\\
$t_R$ & $\frac{1}{6}g_3^2+\frac{1}{18}g_1^2+\frac{1}{8}y_t^2$\\
$b_R$ & $\frac{1}{6}g_3^2+\frac{1}{18}g_1^2+\frac{1}{8}y_b^2$\\
 $H_1$ &  $\frac{3}{16}g_2^2+\frac{1}{16}g_1^2+\frac{1}{4}y_b^2+\frac{\lambda_1}{4}+\frac{\lambda_3}{6}+\frac{\lambda_4}{12}$\\
 $H_2$  & $\frac{3}{16}g_2^2+\frac{1}{16}g_1^2+\frac{1}{4}y_t^2+\frac{\lambda_2}{4}+\frac{\lambda_3}{6}+\frac{\lambda_4}{12}$\\
 $\Sigma$ & $\frac{g_2^2}{2}+\frac{5}{12}b_4$\\ 
 $S$ & $\frac{1}{4}b_4^{(S)}$ 
\end{tabular}
\end{ruledtabular}
\end{table}
\\
\\
\section{Analytical integration for the boundary problem}\label{sec:appendixBaryogenesis}
Solving the boundary problem for the densities varying across the bubble wall is equivalent to considering $N$ coupled linear second-order differential equations
\begin{eqnarray}
y_i''(z)+a_i\,y'_{i}(z)+b_{ij}(z)y_j(z)=s_i(z),\label{eq:coupledset0}
\end{eqnarray}
where index $i=1,\dots, N$ is fixed and not summed over. For our problem $a_i\sim v_w/D_{i}$ are constants while $b_{ij}(z)\sim \Gamma_{ij}(z)/D_{i}$ vary across the bubble wall and differ in the symmetric and the broken phases. However away from the bubble wall $|z|\gg L_w$ the functions $b_{ij}(z)$ converge to constant numbers which we define as: for $z>0,$\,\,$b_{ij}(z)\approx b^r_{ij}$ and for $z<0,$\,\,  $b_{ij}(z)\approx b^l_{ij}$\,, where superscripts $r,l$ stand for ``right" and ``left" corresponding to the ``broken" and ``symmetric" phases respectively.

It is convenient to reduce the system of equations \eq{eq:coupledset0} to the $2N \times 2N$ {\it{first order}} coupled differential equations by introducing a new variable
\begin{eqnarray}
Y(z)\equiv\left( \begin{array}{c}
y_i(z)\\
y'_i(z)  \end{array} \right)\,,
\end{eqnarray}
which is a column vector of the size $2N$\,. In the column notation the set  \eq{eq:coupledset0} reduces to
\begin{eqnarray}
Y'(z)=A\,\mcdot Y(z)+S(z)\,.
\end{eqnarray}
In the equation above $A$ is a $2N\times 2N$ dimensional matrix and $S$ is a $2N$ dimensional column vector which in the block-diagonal form equal to
\begin{eqnarray}
A=\left( \begin{array}{cc}
0 &         \mathds{1}\\
-a & -b  \end{array} \right),\qquad S(z)=\left( \begin{array}{c}
0\\
 s(z)\end{array} \right)\,,
\end{eqnarray}
where $a$ and $b$ are both $N\times N$ dimensional matrices with $a\equiv \diag\,{a_i}$ and  $b$ matrix has matrix elements equal to $b_{ij}$ and we emphasized again that we are working under the assumption of no $z$ dependence of the matrix $A$\,. $N-$dimensional column vector $s(z)$ has elements equal to $s_i(z)$\,. Define $U$ to be a matrix that has its columns consisting of the eigenvectors of the matrix $A$ and assume that it diagonalizes the matrix $A$ according to $U^{-1}A U=A_{\text{diag}}=\diag\,\lambda_k$, where $\lambda_k$ are the eigenvalues of the matrix $A$\,. Because we assume that $A$ has no $z-$ dependence, likewise we obtain that $U, \lambda_k$ have no $z-$ dependence. In this case the transformation $\tilde{Y}=U^{-1}Y$ leads to a simple {\it{uncoupled}} set of $2N$ linear differential equations of the first order
\begin{eqnarray}
\tilde{Y}'_{k}=\lambda_k\,\tilde{Y}_{k}+\tilde{S}_k(z),\qquad \tilde{S}=U^{-1}S(z)\,.
\end{eqnarray}
The solution in terms of the initial conditions for $\tilde{Y}$ is
\begin{eqnarray}
\tilde{Y}_k(z)=\e^{\lambda_k z}\left[\tilde{Y}_k(0)+\int_0^{z}\,d t\,\e^{-\lambda_k t}\,\tilde{S}_k(t)\right]\,.
\end{eqnarray}
Note that in the symmetric phase ($z<0$) and in the broken phase ($z>0$), the matrices $U, \lambda_k, \tilde{S}$ are different\,. We will assume a superscript $l, r$ where appropriate to identify on which side of the bubble wall we are studying the solution to the boundary problem. The boundary problem requires $Y(-\infty)=Y({+\infty})=0$ which easily translates into the boundary condition for the rotated variables $\tilde{Y}(-\infty)=\tilde{Y}({+\infty})=0$\,. Note that while the formulation of the boundary problem requires that the column vector $Y(z)$ is continuous across the bubble wall $Y(-0)=Y({+0})$, the same is not true for $\tilde{Y}(z)$: $\tilde{Y}(-0)~\ne~\tilde{Y}({+0})$. The necessary conditions for the boundary problem to have a solution is 
\begin{eqnarray}
&&\tilde{Y}_k(+0)=-\int_0^{\infty}\,d t\,\e^{-\lambda^r_k t}\,\tilde{S}_k(t),\qquad \lambda^r_k\ge0,\label{eq:necessaryconditions1}\\
&&\tilde{Y}_k(-0)=-\int_0^{-\infty}\,d t\,\e^{-\lambda^l_k t}\,\tilde{S}_k(t),\,\,\,\,\,\,\,\,\, \lambda^l_k\le0\,,\label{eq:necessaryconditions2}\\
&&\lim_{z\rightarrow+\infty}\int_0^{z}d t\,\e^{\lambda^r_k(z-t)}\,\tilde{S}_k(t)=0, \,\,\,\,\,\,\,\,\,\,\,\lambda_{k}^r<0,\label{eq:necessaryconditions3}\\
&&\lim_{z\rightarrow-\infty}\int_0^{z}d t\,\e^{\lambda^l_k(z-t)}\,\tilde{S}_k(t)=0, \,\,\,\,\,\,\,\,\,\,\,\lambda_k^l>0\,.\label{eq:necessaryconditions4}
\end{eqnarray}
Note that all integrals in equations \eq{eq:necessaryconditions1}-\eq{eq:necessaryconditions4}\, are finite which follows from integrability of $\tilde{S}_{k}(t)$ near $\infty$\,. Simultaneous solution of the equations above together with continuity condition at $z=0$ solves the boundary problem. If such solution does not exist then the boundary problem has no solution.

In practical applications it is often used the following VEV profile functions
\begin{eqnarray}
&&\tilde{v}(z)=\frac{1+\tanh\frac{z}{L_w}}{2}\,,\,\,\,\tilde{\beta}'(z)=\frac{1}{2L_w\cosh^2\frac{z}{L_w}},
\end{eqnarray}
with $s_i\sim \tilde{v}(z)^\alpha\tilde{\beta}'(z)^{\beta}$\,. For example in the SUSY example $\alpha=2, \beta=1$. In the present paper we have $\alpha=4, \beta=1$\,.
The following master integral we evaluate analytically
\begin{small}
\begin{eqnarray}
&&I(z,\lambda;\alpha,\beta)=\int_0^z\,d t\,\e^{-\lambda t}\,\left[\tilde{v}(t)\right]^{\alpha}\,\left[\tilde{\beta}'(t)\right]^{\beta}\nonumber\\
&&=\frac{2^{\beta}L_w^{1-\beta}}{2(\alpha+\beta)-L_w\lambda}\e^{\frac{2(\alpha+\beta)t}{L_w}-\lambda t}\\
&&\times_2 F_1\left(\alpha+2\beta,\alpha+\beta-\frac{L_w\lambda}{2},1+\alpha+\beta-\frac{L_w\lambda}{2};-\e^{\frac{2t}{L_w}}\right)\Bigg |_{0}^{z}.\nonumber
\end{eqnarray}
\end{small}
Consistently with our assumption that the relaxation rates are approximately given by a step function across the bubble wall it is safe to assume that $|z|\gg L_w$ in the equation above. Indeed the BAU is generated far away from the bubble wall and this is an extremely reliable approximation. Thus by taking limits of the hypergeometric function at infinity obtain
\begin{small}
\begin{eqnarray}
&&I(+\infty,\lambda;\alpha,\beta)\nonumber\\
&&=\frac{2^{\beta}L_w^{1-\beta}}{2(\alpha+\beta)-L_w\lambda}\Bigg[\frac{\Gamma\left(1+\alpha+\beta-\frac{L_w \lambda}{2}\right)\Gamma\left(\beta+\frac{L_w \lambda}{2}\right)}{\Gamma\left(\alpha+2\beta\right)}\nonumber\\
&&-_2 F_1\left(\alpha+2\beta,\alpha+\beta-\frac{L_w\lambda}{2},1+\alpha+\beta-\frac{L_w\lambda}{2};-1\right)\Bigg]\nonumber\\
&&\qquad\qquad\qquad\qquad\qquad\qquad\text{for}\,\,\lambda > -\frac{2\beta}{L_w}\,.\label{eq:plusinfinitylimit}
\end{eqnarray}
\end{small}
In the opposite limit $z\rightarrow -\infty$ we find 
\begin{small}
\begin{eqnarray}
&&I(-\infty,\lambda;\alpha,\beta)\nonumber\\
&&=-\frac{2^{\beta}L_w^{1-\beta}}{2(\alpha+\beta)-L_w\lambda}\label{eq:minusinfinitylimit}\\
&&\times_2 F_1\left(\alpha+2\beta,\alpha+\beta-\frac{L_w\lambda}{2},1+\alpha+\beta-\frac{L_w\lambda}{2};-1\right),\nonumber\\
&&\qquad\qquad\qquad\qquad\qquad\qquad\text{for}\,\,\lambda < \frac{2(\alpha+\beta)}{L_w}\,.\nonumber
\end{eqnarray}
\end{small}
The equations \eq{eq:plusinfinitylimit} and \eq{eq:minusinfinitylimit} are directly applicable in \eq{eq:necessaryconditions1} and \eq{eq:necessaryconditions2}\,. Finally we also establish the following limits 
\begin{eqnarray}
&&\lim_{z\rightarrow+\infty} \,\e^{\lambda z}\,I(z,\lambda;\alpha,\beta)=0, \qquad \text{for} \,\,\lambda<0,\\
&&\lim_{z\rightarrow-\infty} \,\e^{\lambda z}\,I(z,\lambda;\alpha,\beta)=0, \qquad \text{for} \,\,\lambda>0\,.
\end{eqnarray}

The equations above demonstrate that \eq{eq:necessaryconditions3} and \eq{eq:necessaryconditions4} automatically are satisfied.

\vskip 0.1in

\bibliographystyle{h-physrev}
\bibliography{bibliography}

\end{document}